# Rubin ToO 2024:
# Envisioning the Vera C. Rubin Observatory LSST Target of Opportunity program

*Authors and Lead Editors:* Igor Andreoni (1,2,3,4), Raffaella Margutti (5,6)

*Authors and Section Editors:* John Banovetz (7), Sarah Greenstreet (8,9), Claire-Alice Hébert (7), Tim Lister (10), Antonella Palmese (11), Silvia Piranomonte (12), S. J. Smartt (13,14), Graham P. Smith (15), Robert Stein (16)

*Authors and endorsers:* Tomas Ahumada (17), Shreya Anand (18,19,20), Katie Auchettl (21,22), Michele T. Bannister (23), Eric C. Bellm (9), Joshua S. Bloom (5,24), Bryce T. Bolin (25), Clecio R. Bom (26,27), Daniel Brethauer (5), Melissa J. Brucker (28), David A.H. Buckley (29,30,31), Poonam Chandra (32), Ryan Chornock (5), Eric Christensen (8), Jeff Cooke (33,34), Alessandra Corsi (35), Michael W. Coughlin (36), Bolivia Cuevas-Otahola (37), D'Ammando Filippo (38), Biwei Dai (6,24), S. Dhawan (39), Alexei V. Filippenko (5), Ryan J. Foley (22), Anna Franckowiak (40), Andreja Gomboc (41), Benjamin P. Gompertz (15,42), Leanne P. Guy (8), Nandini Hazra (43,44,45), Christopher Hernandez (46), Griffin Hosseinzadeh (47), Maryam Hussaini (48), Dina Ibrahimzade (5), Luca Izzo (49,50), R. Lynne Jones (8), Yijung Kang (19,20), Mansi M. Kasliwal (16), Matthew Knight (51), Keerthi Kunnumkai (11), Gavin P Lamb (52), Natalie LeBaron (5), Cassandra Lejoly (28), Andrew J. Levan (53,54), Sean MacBride (55), Franco Mallia (56), Alex I. Malz (57), Adam A. Miller (58,59), John Carlos Mora (J. C. Mora) (60,61,62), Gautham Narayan (63,64), Nayana A.J. (5), Matt Nicholl (65), Tiffany Nichols (66,67,68), S. R. Oates (69), Akshay Panayada (70), Fabio Ragosta (71,72), Tiago Ribeiro (8), Dan Ryczanowski (73,15), Nikhil Sarin (74,75), Megan E. Schwamb (76), Huei Sears (77), Darryl Z. Seligman (78,79), Ritwik Sharma (80), Manisha Shrestha (81), Simran Kaur (82), Michael C. Stroh (59), Giacomo Terreran (10), Aishwarya Linesh Thakur (83), Aum Trivedi (84), J. Anthony Tyson (85), Yousuke Utsumi (86), Aprajita Verma (87), V. Ashley Villar (48,88), Kathryn Volk (89), Meet J. Vyas (84), Amanda R. Wasserman (63,64), J. Craig Wheeler (90), Peter Yoachim (9), Angela Zegarelli (40)

*Author and SCOC Chair:* Federica Bianco (91,92,93,94)

Affiliations can be found at the end of the document.

# Introduction

The Legacy Survey of Space and Time (LSST) at Vera C. Rubin Observatory is set to start in late 2025. The LSST survey cadence is designed in an iterative community-driven process (Bianco et al., 2022). The implementation of a Target of Opportunity (ToO) program using up to 3% of the LSST sky time has been recommended by the Survey Cadence Optimization Committee (SCOC; see PSTN-055[1]). This program is recommended to initially respond to Gravitational Wave triggers but with the potential to expand to other phenomena of interest.

A workshop has been organized to bring together experts from the scientific community, Rubin Observatory personnel, and members of the Rubin SCOC to deliver a recommendation for the implementation of the ToO program. The workshop took place on March 18-20, 2024 at the University of California, Berkeley. It was designed with plenary discussion to explain the boundary conditions for ToOs set by the Observatory, and collaborative breakouts to define the needs of identified scientific cases and develop trigger and prioritization schemes.

The SCOC, with the support of Rubin Observatory, has prepared a template for this report to collect a consensus recommendation.

Four main science cases were identified: gravitational waves (GWs) multi-messenger astronomy, high energy neutrinos, Galactic supernovae, and small potentially hazardous asteroids (PHA) potential impactors. The GW multi-messenger astronomy science case was analyzed differently for various types of GW sources: binary neutron star and neutron star–black hole mergers, gravitationally lensed binary neutron star mergers, binary black hole mergers, and unidentified GW sources. Additional science cases were identified, among which gravitationally lensed gamma-ray bursts and their afterglows, and future interesting twilight discoveries have been briefly addressed in this document. Follow-up of poorly localized gamma-ray bursts is also among the science cases that should be studied in the future.

For each science case, we discussed how to select and prioritize triggers, automate response to triggers, and what exposures, filters, and cadence are required. This document represents the product of the Rubin ToO 2024 workshop[2], with additional contributions from members of the Rubin Science Collaborations.

As part of this recommendation, we recommend that the Rubin ToO program be revised yearly by the SCOC with input from the community. During the International Gravitational-Wave Observatory Network (IGWN) observing runs, in which the LIGO/Virgo/KAGRA detectors are operational, we suggest that a revision of the ToO program takes place every 6 months.

On October 1, 2024, the SCOC released the "Survey Cadence Optimization Committee's Phase 3 Recommendations" document (PSTN-056[3]). There, it is recommended "the implementation of a LSST ToO program as detailed in the community report `Rubin ToO 2024: Envisioning the Vera C. Rubin Observatory LSST Target of Opportunity program` by the scientific community at large'." Although a final decision is still pending, the following document can be considered as a reasonable baseline plan for ToO observations with the Rubin Observatory.

[1] https://pstn-055.lsst.io
[2] https://lsst-tvssc.github.io
[3] https://pstn-056.lsst.io/

# Table of Contents

# Science Case: Gravitational Waves and Multi-Messenger Astronomy

In this scientific context, we outline our recommendation for the allocation of ToO time suggested to respond to gravitational waves (GW) and multi-messenger astronomy (MMA) triggers, delineating various types of scientific targets pertinent to this field as the binary neutron star mergers and neutron star - black hole mergers (BNS, NS-BH), gravitationally lensed binary neutron star mergers, black hole-black hole mergers (BBH) and unidentified GW sources.

## Science case justification and description: Binary Neutron Star Mergers and Neutron Star - Black Hole Mergers

*Science case leads: Silvia Piranomonte, Stephen Smartt*

Detecting the electromagnetic (EM) counterparts of gravitational wave sources provides a wide range of science opportunities as demonstrated by the remarkable discovery of GW170817, GRB170817A, and AT2017gfo due to a binary neutron star (BNS) merger. This can illuminate the origin of the r-process elements and determine if compact binary mergers are either the sole or dominant source of elements heavier than iron. The maximum mass of neutron stars and the neutron star equation of state can be probed along with the nature of objects that are known to sit in the “mass-gap” between neutron stars and black holes. Electromagnetic detections are essential to exploit the full opportunity of gravitational wave physics, and Rubin has the unique potential to revolutionize the field. We project that over the 10 years of Rubin, we have the potential to detect 50 counterparts to gravitational wave triggers (BNS or Neutron Star-Black Hole, NS-BH), which would allow a 2% measurement of the Hubble constant $H_0$. Open questions in the field such as the link between both short and long gamma-ray bursts and compact binary mergers and tests of the endpoints of the supernova mechanism (through binary stellar population synthesis) can be addressed with a large sample. We expect kilonovae from BNS and NS-BH mergers to have a wide range of ejecta masses and compositions, ranging from the light r-process elements to the 3rd peak dominated by Pt, Au, and Os. These may appear quite different, given the high opacity of the lanthanides, actinides, and 3rd peak elements due to the multitude of bound-bound transitions in these electronically complex ions. While a sample of 50 is an ambitious plan over the lifetime of the Legacy Survey of Space and Time (LSST), we focus our attention on what we can achieve during the LIGO-Virgo-KAGRA (LVK) Observing Run 5 (O5) which is projected to begin in June 2027.[4]

### Justification for the use of Rubin

Rubin is the only facility capable of covering the skymaps produced by LVK (expected to be in the region of 100-500 square degrees during O5) and at the same time reaching the depths required for kilonova detection and recognition at distances of 350-700 Mpc (see Table 1). During 2025-2035, when LVK reaches its design sensitivity, there will be no other facility that has the ability to reach these distances consistently. A comprehensive discussion of potential strategies compared to example lightcurves of kilonovae out to 300 Mpc was published in

[4] https://observing.docs.ligo.org/plan

Andreoni et al., 2022 (which evolved from the TVS Science Collaboration whitepaper), and here we build on that with updates from the latest information and projections for O5.

| Filter | Depth (AB mag) | *M* (350 Mpc) | *M* (700 Mpc) | Exptime (sec) |
|---|---|---|---|---|
| *u* | 24.9 24.7 23.9 | -12.8 -13.1 -13.8 | -14.4 -14.6 -15.3 | 180 - 120 - 30 |
| *g* | 26.0 25.8 25.0 | -11.7 -12.0 -12.7 | -13.3 -13.5 -14.2 | 180 - 120 - 30 |
| *r* | 25.7 25.5 24.7 | -12.0 -12.3 -13.0 | -13.6 -13.8 -14.5 | 180 - 120 - 30 |
| *i* | 25.0 24.8 24.0 | -12.7 -13.0 -13.7 | -14.3 -14.5 -15.2 | 180 - 120 - 30 |
| *z* | 24.3 24.1 23.3 | -13.4 -13.7 -14.6 | -15.0 -15.2 -15.9 | 180 - 120 - 30 |
| *y* | 23.1 22.9 22.1 | -14.6 -14.9 -15.6 | -16.2 -16.4 -17.1 | 180 - 120 - 30 |

**Table 1:** The 30-second 5-sigma depths are taken from Bianco et al., 2022 (in orange). The absolute magnitudes in each filter are given at two reference distances (350 Mpc and 700 Mpc). These depths are scaled to 120s and 180s assuming we are background limited. Relevant exposure times are provided for those absolute magnitudes.

## Rates and Target of Opportunity activation criteria

**Event rate (triggers/year):** 16+6 during O5

**Trigger time distribution or constraints:** All during O5.

**Trigger format:** The content of the packets is included in Appendix A Table A1.

The major uncertainties in planning the Rubin ToO program are the rate of BNS mergers ($R_{BNS}$) and the size of the sky localization maps. We have employed the official LVK observing scenarios study (Kiendrebeogo et al., 2023) which assumes an (optimistic) HLVK detector configuration[5]. If we select only GW events with high signal-to-noise (SNR > 12) then the projected numbers of events per year with the O5 sensitivity are given in Table 2. The projections for O5 are 37 BNS with 90% probability sky localization areas $\Omega < 1000$ square degrees, of which 32 have $\Omega < 500$ etc.

While there is some uncertainty in the above, the naive estimate of numbers of mergers based on the published volumetric rate of $R_{BNS} = 98^{+260}_{-85}$ Gpc$^{-3}$ yr$^{-1}$ (90% CL; Abbott et al., 2023a; Table 2 Binned Gaussian Probability, BGP, model) and a BNS range of the LIGO detectors of 280 ± 40 Mpc indicates approximately $R_{BNS} = 19^{+33}_{-18}$ yr$^{-1}$ BNS detections per year. This estimate broadly agrees, within a factor of 2 with Table 2.

Considering $R_{NS\text{-}BH} = 32^{+62}_{-24}$ Gpc$^{-3}$ yr$^{-1}$ (Abbott et al., 2023a; Table 2 BGP model, Colombo et al., 2023) estimate around 10 NS-BH per year within 300 Mpc, again in reasonable agreement with Table 2.

We assume that O5 begins in June 2027 and will run for 2.5 years until the end of 2029. We also assume the LSST begins in October 2025.

We will only recommend triggering on high-significance BNS or NS-BH events with the following high-cadence strategy (lensed BNS, unmodelled bursts, and BBH are discussed separately). We consider the False Alarm Rate (FAR) provided by LVK as the primary source of reliability and require the following parameters to be set in the GW Kafka alert packet[6] :

[5] https://emfollow.docs.ligo.org/userguide/capabilities.html

### Trigger Criteria

- *Only trigger on an Initial map, do not trigger on Preliminary*[7]
- *The probability of being BNS or NS-BH should be greater than 90%:* ***BNS+NS-BH >= 0.9***
- *False alarm rate less than 1 per 1 year:* ***FAR < 1.6e-08 Hz***
- *90% sky area less than 500 square degrees (then we consider one of the two scenarios below):* ***Ω < 500***
- *For NS-BH events, require that there is a good probability that mass has been ejected (these numbers will be changed based on O4 results and O5 projections):* ***HasNS >= 0.5 and HasRemnant >= 0.5,*** *which correspond to the probability that at least one of the two merging objects is a neutron star and the probability that the system ejected a non-zero amount of neutron star matter*[8]*, respectively.*

## Time budget and observing strategy

| Sky area $\Omega$ (sq deg) | Median number of BNS | 90% confidence range of BNS with maps < $\Omega$/year | 90% confidence range of NS-BH with maps < $\Omega$/year |
|---|---|---|---|
| 50 | 14 | 6-31 | 0 - 3 |
| 100 | 18 | 8-41 | 1 - 4 |
| 150 | 20 | 9-45 | 1 - 5 |
| 250 | 25 | 11-56 | 1 - 7 |
| 500 | 32 | 14-70 | 2 - 9 |
| 1000 | 37 | 17-83 | 2 - 11 |

**Table 2**: Projected ranges of the number of BNS and NS-BH mergers in O5 (90% confidence) with the expected sky map size, located anywhere in the sky. These values must be corrected for observability constraints by a factor of ~1/3 for Rubin. The ranges highlight the current uncertainty in rates and projected detector performance in O5. In the rest of this document, we base our time ToO recommendation on the median number from the calculation described in the text.

To estimate the telescope time required to cover each of these sky areas, we assume that Rubin has an effective FoV of 7 square degrees rather than 10 square degrees as some overlap between the footprints will be required. We also assume 4.8 seconds slew and readout time and 120 seconds for a filter change. We assume 60% of the sky has been covered by the WFD footprint

[6] These trigger criteria are based on what we currently see in O4 and will be updated for O5.

[7] A *Preliminary* map is automated, an *Initial* map has been human vetted. The time difference is typically minutes.

[8] https://emfollow.docs.ligo.org/userguide/content.html

(for templates) and a further 65% is available in the nighttime sky hence roughly ⅓ of the events in Table 2 are available for immediate Rubin observing.

Taking into account several kilonova lightcurves simulations which provide both "bright" and "faint" sources at two different distances, 350 Mpc and 700 Mpc (Figure 1), we propose the following observing strategies for the joint BNS+NS-BH science case distinguishing "gold" and "silver" events (see also Figure 2):

1. Gold events: small skymap (100 sq deg) and FAR < 1 $yr^{-1}$

For gold events, we mean the case in which we will have a small skymap (100 sq deg) and FAR < 1 $yr^{-1}$. In terms of observability constraints, during O5 we expect to trigger only 1/3rd of the 18 triggers localized better than 100 sq deg (see Table 2) which means *6 BNS* and up to *2 NS-BH events per year with* $\Omega$ *< 100* square degrees.

We might then have the following observing scenario which we refer to as "3-filter deep strategy" (night 0 is defined as the first period of nighttime hours after the GW trigger):

- **Night 0**: three scans across the skymap in 3 filters with 120-sec exposure for each filter *gri*[9]. Each scan is estimated to take ~95 minutes.
- **Night 1,2,3**: one scan across the skymap in two filters (likely *r*+*i*) with 180 seconds per visit. Each night is estimated to take ~92 minutes.

If the counterpart is not identified by Night 2, we recommend repeating observations on a fourth night (Night 3). Therefore, we assume that 4 epochs are required.

Over the 10-year lifetime of LSST, our target of 50 counterparts may require around 60-70 triggers of this type, depending on the LVK operational plan post O5. This projects to *a maximum of 630 hrs*, which is close to the 3% survey time available. We note that this calculation assumes we always require 4 epochs to complete the identification, which is unlikely in all cases. It also assumes we will always have a Night 0 in which we will trigger three scans in 3 filters. Nevertheless, in many cases, the time elapsed between the detection of gravitational waves and the accessibility of the skymap will lead us to move directly to the Night 1 case (one scan) and proceed from there.

Night 0 timing considerations: we calculate that it will take 95 minutes (1.59 hrs) to scan a 100 sq deg skymap and the scheduling of the three epochs on Night 0 will depend on the observability window of the sky map ($T_{win}$). Here we provide a timing strategy:

➢ If $T_{win}$ >=3*1.59 = 4.76 hrs; do all 3 scans, with the gap between them of ($T_{win}$ - 3*1.59)/2)
➢ Else if 3.2 < $T_{win}$ < 4.7 hrs; do 2 scans. The start time of the visits is separated by ($T_{win}$ - 2*1.59), which is sufficient for asteroid rejection and short timescale evolution.
➢ Else if $T_{win}$ < 3.2 hrs then do 1x1.59 hrs scan + followed by 1 scan (0.53 hrs) in *r*-band only after the full filter complement is complete, to ensure asteroid rejection is reliable.

[9] Depending on the results of kilonova sensitivity simulations, we may switch this to *grz*.

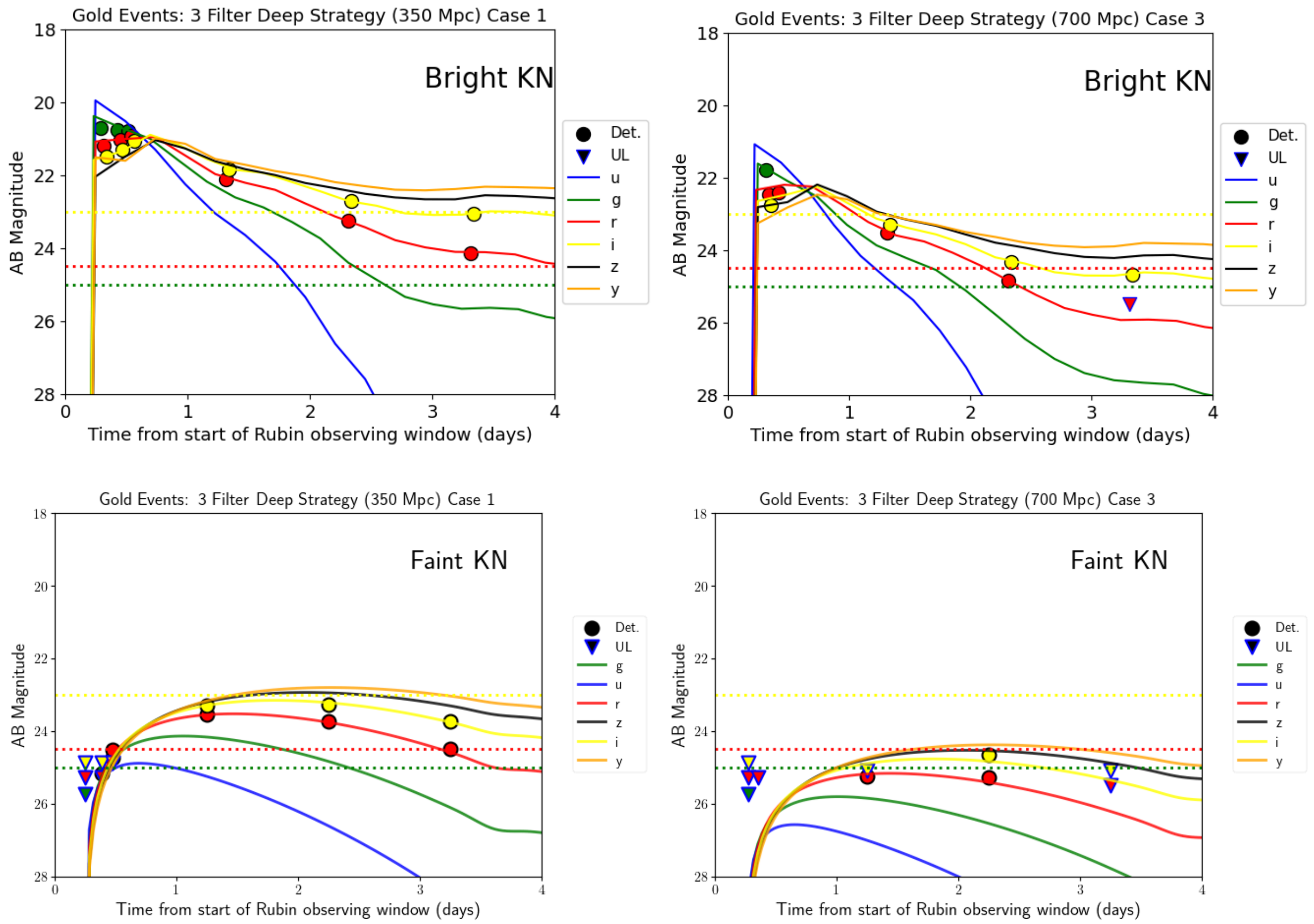

**Figure 1**: Example of kilonova (KN) lightcurves simulations illustrating two suites of physically plausible models that provide "Bright" (top panels) and "Faint" (bottom panels) sources. The Bright kilonova models are taken from Bulla (2019) and parameters as in Anand et al., (2024). The Faint kilonova models are from Sarin et al., (2023). Deep 120 sec exposures are employed. The distances chosen are 350 Mpc (left panels), which is likely the typical distance for BNS detections in O5, and twice the distance which is possible if the inclination angle is favorable and the masses at the upper end of the NS mass distribution (700 Mpc, right panels). Given the uncertainties in the luminosity function of kilonovae, we chose the 3-filter deep strategy. The bottom right panel shows that even with the deep strategy we may not be complete to the faintest kilonovae at the largest possible distance.

Within the 16 triggers for $\Omega < 100$ sq deg, we may expect to (optimistically) find a number of extremely well-localized ($\Omega \sim 30$ sq deg; similar to AT2017gfo, Abbott et al., 2017). In these cases, we would extend the 3-filter strategy to a 5-filter strategy on Night 0, but keep well within the 9 hrs per trigger. For example, a 30 sq deg skymap requires ~4 hrs to cover Night 1 in *ugriz* (120 sec per visit) keeping the remaining nights' strategies the same.

Gold event summary: *expected triggers during O5 (covering Rubin operation years 2025.8 - 2030.0) = 16; A total of 9 hrs per trigger is recommended =>* ***144 hrs which would be triggered during the O5 run***.

## 2. Silver events : medium skymap events (500 sq deg) and FAR < 1 $yr^{-1}$

Silver events are defined as $100 < \Omega < 500$ sq deg and FAR < 1 $yr^{-1}$. Based on Table 2, in O5 we may expect 4 BNS per year and 2 NS-BH per year with $\Omega < 500$ sq deg skymaps. We will employ a 3-filter strategy with only 1 scan over the skymap per night in two filters.

We might then have the following observing scenario:

- **Night 0**: one scan across the 500 sq deg skymap with 2 filters, 30-sec exposure on each filter (*g+i* or *g+z)*. Each scan is estimated to require ~1.45 hrs.
- **Night 1,2,3**: one scan across the skymap in the same two filters at 120 seconds per visit which will take 5.02 hrs on each night.

If the counterpart is not identified by Night 2, we will repeat observations on a fourth night (Night 3). We will assume that 4 epochs are required.

Silver event summary: *expected triggers during O5 (covering Rubin operation years 2025.8 - 2030.0) = 6. A total of 16 hrs per trigger is recommended => **96 hrs during the O5 run**.*

### 3. Very large sky maps

It is possible we will encounter highly significant events that have a high probability of being a BNS or NS-BH (BNS+NS-BH >= 0.9) but are poorly localized. This is most likely due to detector duty cycles as in the case of GW190425. In such cases where the skymap has $\Omega > 1000$ square degrees, we recommend that survey simulations are run to validate the possibility of adjusting the LSST WFD survey strategy to cover the skymap with standard survey observations. We would be recommending an adjustment to the upcoming observing sequence over the lunation such that the GW skymap is prioritized over 3 nights. We would impose a maximum of 4 of these interventions in O5 and are willing to work with the Rubin Observing strategy team and SCOC to achieve minimal impact on Wide Fast Deep (WFD) strategy.

## Total BNS+NS-BH summary recommendation

| Type | Skymap size | Number of triggers | Time per trigger | Total |
|---|---|---|---|---|
| Gold (3-filter+deep) | 100 | 16 | 9 hrs | 144 hrs |
| Silver | 500 | 6 | 16 hrs | 96 hrs |
| **Grand total** | | | | **240 hrs** |

**Table 3:** Total Summary recommendation for BNS and NS-BH during the LVK O5 run

The recommended 240 hrs will likely span the timeframe from the start of LSST through the LVK O5 observing run (October 2025 to end of 2029) or roughly 4 years. The estimates of 3% of Rubin are between 50 - 60 hours per year or in the region of 200 - 240 hrs available for ToO over this period.

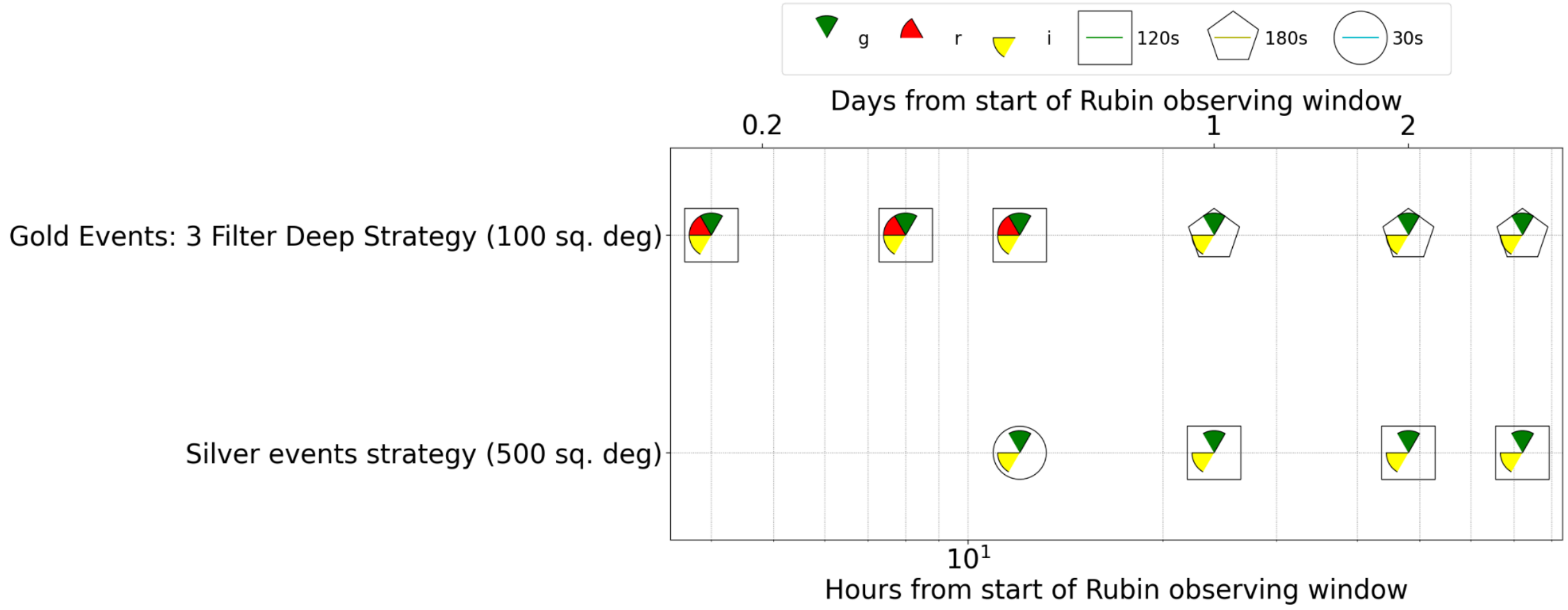

**Figure 2**: Schematic diagram for the proposed strategies for the joint BNS+NS-BH science case. Gold Events are small skymap events (100 sq deg) with FAR<1 per year, to be followed up with a 3-filter deep strategy. Silver events are medium skymap events (500 sq deg) and FAR < 1 per year.

## Downscoping direction suggestions

The descope options are simply to trigger on fewer events; we prefer to retain the strategy set out here and observe fewer events than reduce the filter-area-exposure time

## Non-standard observing and data processing requests

We have recommended 120s and 180s exposures. These can either be 4x30s (6x30s) or 1x120s (1x180s) and should be chosen in consultation with the Rubin Data Management (DM) team and the camera team. The 1x120s (1x180s) implementation would not put any further stress on the standard nightly processing but saturation effects (including bleeds, cross-talk, and persistence) require investigation in LSSTCam. The 4x30s (6x30s) strategy would require those to be stacked before differencing, which would be non-standard processing.

## Science case justification and description: Gravitationally lensed Binary Neutron Star mergers

*Science case lead: Graham Smith*

The first secure and unambiguous detection of a gravitationally lensed merger of a binary compact object will be a major milestone in modern science that unlocks a wide range of novel and high-impact science. Crucially in the context of Rubin/LSST ToO's, a single robust discovery will serve as a "golden object" analogous to the Hulse Taylor pulsar and GRB170817A / GW170817 / AT2017gfo. Achieving this breakthrough requires the detection of multiple messengers from the lensed binary compact object merger because the combination of multiple messengers *significantly* enriches the science and electromagnetic (EM) detection is the *only proven way* to localize a gravitational wave (GW) source to sub-arcsecond uncertainty. Rubin/LSSTCam is uniquely powerful for this science case, offering the only realistic prospect for overcoming the localization challenge, thanks to their unique combination of collecting area, field-of-view, and fast slew speed. We propose Target of Opportunity (ToO) observations of up to one candidate lensed binary neutron star (BNS) merger per year, carefully selected from human-vetted LIGO-Virgo-KAGRA (LVK) alerts. The ToO objective is to detect a candidate lensed kilonova counterpart that we will then confirm with space-based follow-up. The typical lensed BNS merger is predicted to be at redshift z≈1-2, gravitationally magnified by μ≈100, with lensed images separated by up to one day in the observer's frame (Smith et al., 2023). In addition to shaping our discovery strategy, these properties significantly expand the science beyond fundamental physics (e.g., the first test of GR to combine lensing and GWs) and cosmology (e.g., a new high precision probe of $H_0$) to include new constraints on the existence of a "mass gap" between NS and BH, the NS equation of state from pre-peak rest-frame ultraviolet photometry of the second lensed image, and the first glimpse of EM-bright GW sources at z>1.

### Justification for the use of Rubin

The lightcurves of lensed kilonova counterparts to lensed BNS mergers that are detectable in LVK's O5 are predicted to peak at AB≈23.5-26.5, depending on the assumed kilonova physics, with 170817-like kilonovae being brighter than more conservative models that are redder and viewed at a less favorable angle than AT170817 (Figure 3; Nicholl et al., 2021; Bianconi et al., 2023; Smith et al., 2023). Across all kilonova models, they fade more rapidly than any other extragalactic transient that we are aware of, at >~1 mag per day in Rubin's *gri*-bands (rest-frame ~150-350 nm). Rapid and deep Rubin ToO observations are the *only* way to map hundreds of square degrees (the best-localized candidates, based on current expectations of LVK localizations) to these depths before the putative kilonova counterpart has faded from view for ground-based optical/infrared survey telescopes. We will turn to *JWST* for further photometry and spectroscopic confirmation of credible lensed kilonova candidates detected in the Rubin ToO data because no other facility will be sensitive enough after the first few days post-GW detection (Figure 3).

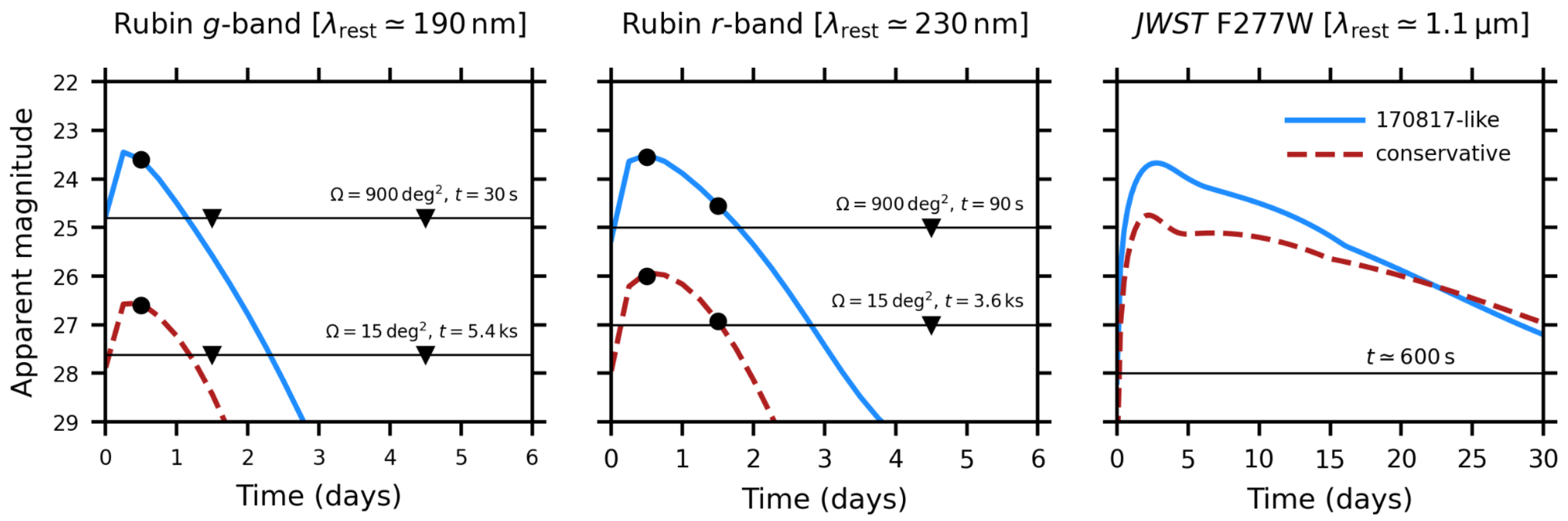

**Figure 3**: Predicted observer-frame light curves of lensed kilonova counterparts to lensed BNS that are detectable in O5. The blue solid curve shows a 170817-like kilonova model and the red dashed curve shows a redder and less favorably inclined kilonova model, based on Nicholl et al., (2021) and Smith et al., (2023). Comparison of the range of apparent magnitude at peak in these lightcurves with that in Figure 1 confirms that the range of kilonova physics explored in this report for BNS that are lensed and not lensed is consistent. The upper horizontal line in the left and center panels shows the threshold at which the proposed Rubin ToOs become capable of probing known kilonova physics – i.e., able to probe deep and wide enough to test the hypothesis of a 170817-like lensed EM counterpart. As the sky localization uncertainties shrink, the proposed ToOs are able to probe deeper (at a fixed total amount of time per epoch) and expand sensitivity to a wider range of kilonova models. For the best localized lensed BNS candidates, we will be sensitive to the conservative model (lower horizontal line). Filled circles are detections and filled triangles are upper limits. As discussed in the text, the sensitivities shown in these panels are constrained by requiring each ToO epoch to last a duration of <~ 5 hours of elapsed time. The right panel shows the predicted infrared lightcurve for the same lensed kilonova counterparts as shown in the other panels – illustrating that they remain detectable with relatively short *JWST* observations for at least a month after the GW detection.

## Rates and Target of Opportunity activation criteria

**Event rate (triggers/year):** <= 1 during O5

**Trigger time distribution or constraints:** All during O5

**Trigger format**: The content of the packets is included in Appendix A Table A1.

Detectable lensed BNS are predicted to be magnified by a large factor, typically $\mu \approx 100$ in O5, and therefore in low latency (when LVK assumes $\mu = 1$) they will appear to be significantly closer and more massive than their true distance and mass. Lensed BNS detections in O5 are predicted to emanate from sources with individual compact object masses in the range ≈ 2 – $10M_\odot$, with ≈ 60% of them located in the "mass gap" between NS and BH – i.e. lensed BNS don't "look like" BNS that are not lensed (Smith et al., 2023). The mass gap is a sparsely populated region of parameter space for GW detections to date. Our approach is therefore to select candidate lensed BNS based on a large value of the *p(HasMassGap)* probability that LVK releases in their alerts. This is the probability that one or more of the merging compact objects has a mass (assuming $\mu = 1$) in the range 3 – 5 $M_\odot$.

Independent of lensing, whether the mass gap is sparsely populated or empty is a very active field of research on the formation of low-mass BH and high-mass NS (e.g., The LIGO Scientific Collaboration et al., 2024). Viewed from that perspective, rapid Rubin ToO follow-up observations of GW detections that LVK identify with low latency as having high

*p(HasMassGap)* probabilities are essential to rule out (at least put limits on) the lensed BNS interpretation of GW detections "in the mass gap".

The rate of detections of lensed BNS mergers by LVK in O5 (true positives) is no more than one per year. More precise statements are not currently possible due to significant uncertainties in the comoving rate density of BNS mergers, and its evolution with redshift. Recent forecasts agree that the rate of lensed detections will be in the range ~0.01–1 per year (Smith et al., 2023; Magare et al., 2023). The same authors show that the relative rate of detections – rate of lensed detections divided by the rate of detections that are not lensed – is much less uncertain, at one lensed detection per thousand detections that are not lensed, with a ≈10% uncertainty (see also Abbott et al., 2023c). Adopting the same official LVK observing scenarios study as discussed in the "BNS mergers and NS-BH mergers" case above (Kiendrebeogo et al., 2023) yields an estimate of 0.04 gravitationally lensed BNS detections per year during O5 with 90% credible sky localizations of $\Omega <= 900$ degree$^2$.

To estimate the rate of false positive detections of GW sources in the mass gap we have simulated the population of BNS, NS-BH, and BBH GW sources that are not lensed that LVK will detect in O5, and also reviewed GW detections in both O3 (Abbott et al., 2023b) and O4 to date. Our simulations use public tools available from LVK[10] and are based on a compact object mass function that is essentially empty in the mass gap. Normalizing the simulations to the same underlying number of detections of BNS that are not lensed in O5 as discussed above in the "BNS and NS-BH mergers" case yields an estimated false positive rate of ~0.2 per year – i.e. 0.2 GW detections per year that would be flagged with *p(HasMassGap)*>=0.9 in O5 that are not lensed. Whilst the uncertainties are large, this implies that the false positives are unlikely to overwhelm the true positives.

In O3 LVK flagged 5 GW detections in their alerts as having *p(HasMassGap)*>=0.9. These sources were later clarified as comprising one NS-BH merger and four BBH mergers with individual BH masses in the range 6–10 $M_\odot$ (Abbott et al., 2021; Bianconi et al., 2023). The NS-BH is clearly not a candidate lensed BNS. In O4 and going forward in O5 it is straightforward to exclude NS-BH from candidate lensed BNS. During O3 LVK included *p(HasMassGap)* in the GW source classification, such that:

$$p(BNS) + p(NS\text{-}BH) + p(HasMassGap) + p(BBH) + p(Terrestrial)=1.$$

In O4, LVK has removed *p(HasMassGap)* from the source classification and now uses it as one of several source properties that inform the EM community of the prospects for EM-bright counterparts. For example, the sole GW source so far in O4 with *p(HasMassGap)*>=0.9 (S230529ay; The LIGO Scientific Collaboration et al., 2024) was classified as *p(NS-BH)*=0.62 and *p(BNS)*=0.31 in its Initial (human-vetted) alert. It could therefore be discarded as a candidate lensed BNS because the source classification provides strong evidence of a mass ratio compatible with being a NS-BH merger. For Rubin ToOs in O5, we therefore intend to suppress contamination by NS-BH that are not lensed by requiring candidate lensed BNS to satisfy both *p(HasMassGap)*>=0.9 and *p(NS-BH)*<0.1. In effect, we will use *p(NS-BH)* as a proxy for the mass ratio of GW sources in LVK alerts.

The four low-mass BBH sources in O3 are bona fide candidate lensed BNS because their masses are within the predicted range for lensed BNS noted above, i.e. ≈ 2 – 10 $M_\odot$. Just two of the four were localized well enough to pass the cut on sky localization uncertainty discussed below ($\Omega$ < 900 degree$^2$), and in line with the "BNS mergers and NS-BH mergers" case, we expect just one-third of GW detections to be actually observable with Rubin. We therefore place a firm upper limit on the number of candidate lensed BNS that may turn out to be low mass BBH that are not lensed of <1 per year.

[10] https://emfollow.docs.ligo.org/userguide/capabilities.html

We will only trigger on high-significance GW detections (judged by LVK's False Alarm Rate, FAR) that pass our selection criteria for candidate lensed BNS events. We require the following parameters to be set in the GW Kafka alert packet:

### Trigger Criteria

- *Only trigger on an Initial human-vetted GW detections*[11]
- *if probability that the GW source includes one or more compact objects in the range 3 – 5* $M_\odot$ *of no less than 90%:* ***p(HasMassGap)>=0.9***
- *if probability that the GW source is a NS-BH merger of less than 10% :* ***p(NS-BH)<0.1***
- *if False Alarm Rate of no more than one per year:* ***FAR<=1.6e-08Hz***
- *if 90% credible GW sky localization of no more than 900 degree*$^2$*:* ***$\Omega$ <=900***

Note that we do not require the detection of two GW signals that are compatible with being lensed images of the same source. However, if two such sources are detected (either within O5, or one in O5 and one in a previous GW run), then we would trigger ToO observations of their joint sky localization if it passes the cut on $\Omega$.

The selection criteria listed above are based on the publicly available information in LVK alerts and are sufficient for the Rubin ToOs described here. Additional selection criteria may be feasible in O5 if LVK is willing to release more information publicly with low latency. Discussions with members of the LVK Lensing Group are ongoing on this topic, and we envisage joint projects to define the most useful information for this purpose following the completion of O4.

We build on the outline observing strategy described by Smith et al., (2023) that is based on the lensed kilonova counterpart being relatively bright and "170817-like" or "conservative", i.e., fainter, redder, and observed at a less favorable viewing angle. The light curves are derived from Nicholl et al., (2021) models. An important feature of this strategy is that the expected number of known strong gravitational lenses will be so high during the Rubin / LSST era (>1 per square degree) that the *only viable strategy is to observe the full sky localization region with Rubin/LSSTCam*. In particular, a strategy based on *pointed follow-up of individual lenses with small field-of-view instruments is not viable*.

## Time budget and observing strategy

**Depth**: Given the faint flux levels and rapid fading, our overall strategy is to go *as deep as possible* over the 90% credible sky localization in the first epoch, which must be feasible within one night of observing time and then repeat twice more. The second is 24 hours after the first, and the third is 2 to 5 days after the second. Epochs two and three aim to detect the rapid fading

[11] A `Preliminary` map is automated, an `Initial` map has been human vetted, the time difference is typically minutes.

(>=1 magnitude per day) and to provide a post hoc template image in which candidate lensed kilonova is predicted not to be detectable, respectively (Figure 3).

**Filters**: We adopt a two-filter strategy, to provide some color information to help constrain models and rule out contaminants, as the first few days after GW detection is the only opportunity to gather optical data. Two filters will also likely assist colleagues with complementary science interests. We select the *g*- and *r*-band filters as they are sensitive and well-matched to the rest frame wavelengths at which the distinctive rapid fade is present. For redder Rubin filters, the reduced sensitivity wins over the slower-evolving light curve, leading to a less efficient observing strategy.

We will also stack all of the *g*/*r*-band data in a single combined *g*+*r* stack per epoch to maximize detection sensitivity, potentially enabling us to push up to ~0.4 magnitudes deeper than the numbers quoted below. Note, that we are not relying on this for the feasibility of the observing strategy.

**Area:** For a representative area of $\boldsymbol{\Omega}$ **= 150 degree²**, we would reach depths of $g \approx 26.2$ and $r \approx 26$. We implement our "*as deep as possible*" approach by adopting a fixed ~5 hours of elapsed time per epoch, with the depth that can be achieved in this time being a function of GW sky localization uncertainty. Actual ToO triggers will fall in the range bracketed by the following two scenarios which are also shown in Figure 3. Note that smaller GW sky localization uncertainties allow us to push to fainter photometric limits and test fainter kilonova models within our fixed envelope of 5 hours per epoch:

> $\boldsymbol{\Omega}$ **= 900 degree²:** This is the threshold at which we can go deep enough in ~5 hours per epoch to test the lensing hypothesis for the AT170817-like model. We estimate that this would require 30 seconds per pointing in *g*-band and 90 seconds per pointing in *r*-band, to reach depths of $g = 24.8$ and $r = 25$.
>
> $\boldsymbol{\Omega}$ **= 15 degree²:** Over this area, we can go deep enough in ~5 hours per epoch to test the lensing hypothesis for a wide range of kilonova models including the conservative model shown in Figure 3. This should require ~1.5 hours of integration time per pointing in *g*-band and 1 hour of integration time per pointing in *r*-band, to reach depths of $g = 27.6$ and $r = 27$.

As a consistency check relative to published photometry of AT2017gfo, we adopt the first *Swift*/UVOT detection (Evans et al., 2017) at AB ≈ -13.5 ($t_{rest} \approx 0.6$ days for a source at effectively z = 0) as the closest match to the first proposed Rubin ToO epoch searching for lensed kilonovae at z > 1 ($t_{obs} \approx 1$ day). Correcting for distance modulus and cosmological k-correction, and applying gravitational magnification of $\mu = 100$ yields an estimated "back of the envelope" magnitude of $g \approx 24 - 25$, i.e., in line with the detailed forecasts summarized in Figure 3 and discussed in detail by Smith et al., (2023).

The proposed observations will uniquely identify a lensed kilonova and distinguish it from more common transient classes, even at these very faint flux levels. The rapid fading of the rest frame ultraviolet kilonova lightcurve alone differentiates it from any type of supernova, "rapidly evolving transient", "fast blue optical transient" (FBOT, e.g., Drout et al., 2014). Flares from cataclysmic variables, stellar flares, and AGN can be ruled out by coincidence with foreground stars and galactic nuclei, respectively. Also, viable lensed kilonovae will be within a few tens of arcsec of a luminous red galaxy, group/cluster of galaxies (e.g., Ryczanowski et al., 2020, 2023) and within a detectable lensed host galaxy. This will allow the absolute magnitude of candidate lensed kilonova counterparts to be estimated when vetting candidates found in the Rubin ToO data, prior to space-based follow-up. By the time that O5 happens, construction of the Rubin/LSST strong lens sample by the Strong Lensing Science Collaboration (SLSC) and Dark Energy Science Collaboration Strong Lensing Topical Team (DESC-SLTT) will also be well advanced, further strengthening the vetting of candidate lensed kilonovae.

Our time recommendation per Rubin ToO trigger is based on the "*as deep as possible in ~5 hours per epoch for 3 epochs*" strategy and is simply ~15 hours. We prefer two triggers, i.e., ~1 per year for the ~2-year duration of O5, because false positives are well controlled but not negligible, as discussed above. This gives a **total recommendation of 31 hours**.

We envisage these 31 hours spanning a period of roughly four years from the start of Rubin operations to the end of 2029. Assuming ≈2000 hours of Rubin on-sky observing per year translates our recommendation of 31 hours to **≈0.4% of the observing time** in these four years, i.e., within the uncertainty on the overall amount of Rubin time devoted to ToO observations.

The ~15 hours per trigger would be spent as follows for the two sky localization sizes that bracket the GW sources that we would follow up (assuming: 4.8 seconds overhead per exposure; 120 seconds per filter change; two filter changes; ~25% overlap between LSSTCam pointings):

**$\Omega$ = 900 degree$^2$:**

- 900 degree$^2$ / 7 degree$^2$ ≈ 128 pointings each in *g*- and *r*-bands
- 5σ sensitivities of $g$ = 24.8 (30 seconds / pointing) and $r$ = 25 (90 seconds / pointing)
- Assume 1 x 30-second exposure in *g*-band and 3 x 30-second exposure in *r*-band

  Time = 3 epochs x 128 pointings x 4 exposures x 34.8 sec + 240 sec = **15.1 hours**

**$\Omega$ = 15deg$^2$:**

- 15 degree$^2$ / 7 degree$^2$ ≈ 2 pointings each in *g*- and *r*-bands
- 5σ sensitivities of $g$ = 27.6 (1.5 hours / pointing) and $r$ = 27 (1 hour / pointing)
- Assume individual exposure times of 180 seconds in both filters
- 2.5 hours / 180 seconds = 50 exposures / pointing

  Time = 3 epochs x 2 pointings x 50 exposures x 184.8 sec + 240 sec = **15.5 hours**

## Downscoping direction suggestions

The main descope option is to restrict the size of the sky localization region on which we would trigger to be smaller than the upper limit of 900 degree$^2$ discussed here. This would reduce the probability of triggering and increase our sensitivity to a broader range of kilonova physics.

We recommend retaining the observing strategy described here because we prefer better-quality data. We also prefer to be able to trigger twice in O5, because it is unlikely that we can control the false positive rate to be negligible.

We also recommend allowing triggering on individual GW detections that we identify as candidate lensed BNS and not to be descoped to only trigger on pairs of GW sources that are consistent with being lensed images of the same object. This is because the latter would compromise both the discovery efficiency (due to finite GW run length and GW detector duty cycle) and the science. In summary, the science would be compromised because the arrival time differences are predicted to be sub-day, therefore rapid follow-up of the first image to arrive is predicted to yield free detection of the rising portion of the rest-frame ultraviolet lightcurve of the second image of the lensed kilonova counterpart. Such a detection would be very powerful for breaking degeneracies in the interpretation of kilonova light curves and constraining the NS equation of state.

## Non-standard observing and data processing requests

Our science goals require deep stacks and inter-night difference imaging. From discussions with Eric Bellm, lead of the Rubin Alert Production (AP) team, we understand that it is possible that the AP team will be able to deliver this data product in time for O5, however, no guarantees are possible currently. Our contingency plan is to build the stacks and do inter-night difference imaging after the 80-hour embargo has elapsed. Whilst sub-optimal, this would not compromise the science, because the primary method for spectroscopic confirmation of candidate lensed kilonova counterparts is with *JWST*/NIRSpec. Lensed kilonovae remain bright enough for such observations for at least a month post-GW detection (Figure 3).

Adjustment of the AP pipeline may be required to enable robust and efficient detection of faint high-magnification lensed image pairs of transients that are not resolved. We are testing this with the AP team as in-kind contributors (SLSC and LSST:UK[12]) to the Difference Image Analysis Science Unit within the Systems Integration, Testing, and Commissioning (SITCom) Rubin team. We will also discuss with the Data Management (DM) and instrument teams the optimal choice of individual exposure times when we all have on-sky experience with LSSTCam.

[12] https://www.lsst.ac.uk

## Science case justification and description: Black Hole-Black Hole Mergers

*Science case lead: Antonella Palmese*

Binary black hole (BBH) mergers occurring in gaseous environments may give rise to an electromagnetic counterpart. In particular, binaries embedded in the disk of an Active Galactic Nucleus (AGN) may be accompanied by electromagnetic (EM) radiation following ram-pressure stripping the Hill sphere around the remnant object (McKernan et al., 2019), accretion (Bartos et al., 2017), shock breakout from jet-disk interaction (Tagawa et al., 2023; Rodriguez-Ramirez et al., 2023), or a jet afterglow. This environment offers an efficient formation channel for stellar-mass binary black hole mergers (McKernan et al., 2012): nuclei of galaxies are expected to host dense populations of black holes, the presence of an AGN disk allows efficient binary formation thanks to migration, and the presence of a deep gravitational potential well allows for efficient retention of kicked remnants, hence it favors hierarchical mergers and can produce high mass stellar-mass BBH and intermediate mass black holes. Population analyses and comparisons with LVK detections show that ~20% (Gayathri et al., 2023) or even up to 80% (Ford et al., 2021) of these BBHs may be explained with an AGN formation channel.

Another intriguing indication in favor of LVK BBH mergers producing EM counterparts in AGN disks, is the sample of candidate counterparts associated with mergers from the previous LVK observing run (Graham et al., 2023), in particular the one associated with GW190521 (Graham et al., 2020), likely a binary composed of two black holes in the upper mass gap. Future GW observing runs overlapping with the lifetime of Rubin LSST will be especially promising in aiding the association of AGN flares with BBH mergers, given that the improved localizations will allow for confident associations given the lower probability of chance coincidence with background flaring AGNs (Palmese et al., 2021).

The identification of a confirmed EM counterpart to a BBH would be a ground-breaking discovery, and enable a wide range of scientific analyses, similar to multi-messenger observations of BNS and NS-BH. These include measurements of the $H_0$, modified gravity constraints (e.g. Belgacem et al., 2018), studies of binary formation mechanisms, and AGN disks (Ford et al., 2019). Even if the GW-AGN flare association cannot be established with confidence for single events, this can be done on a population level to place constraints on the overall contribution of the AGN formation channel to BBH mergers (Palmese et al., 2021) and on $H_0$ (Bom & Palmese 2024). We also note that other models of EM counterparts to BBHs have been proposed (e.g. Perna et al., 2016), e.g. thanks to the presence of a circumbinary disk, and could be characterized with the proposed observations.

Similarly to BNSs, strongly lensed BBHs can also occur (e.g. Hannuksela et al., 2019), and if they arise from AGN disks we may also expect a strongly lensed EM counterpart, enabling similar cosmological measurements to those outlined in the lensed kilonova section (e.g. Hannuksela et al., 2020). The identification of a strongly lensed EM counterpart will also constitute a smoking gun confirming that multiple GW detections are indeed multiple lensed GW images.

## Justification for the use of Rubin

Because only a fraction of BBH mergers are expected to produce AGN flares, and because of the variable nature of AGNs, a systematic follow-up campaign is necessary to achieve the above science goals. Thanks to its combination of wide field and sensitivity, Rubin will be unique in enabling associations between GW mergers and BBH counterparts with a very modest amount of time. Providing estimated luminosities for BBHs in AGN disks is extremely challenging, even more so than for BNS/NS-BH, given the variety of viable emission mechanisms proposed in the various models described above and their strong dependence on the local environment and binary properties. For example, the flare identified in Graham et al., 2020 (peaking at around g~18.8), may have originated from a remnant kick velocity in the order of 200 km/s. Assuming the flare originates as a shocked Bondi tail, simply changing the velocity to that of a strong kick (e.g. 1000 km/s, which is feasible under some configurations e.g. Zlochower & Lousto 2015) and leaving all the other properties the same, would result in a 3 orders of magnitude decrease in luminosity. Considering a BBH with a total mass of 50 $M_\odot$ (the cut we assume) would reduce the luminosity by another order of magnitude, and 75% of the events we select will be at a larger distance than the AGN in Graham et al., 2020. Moreover, it is possible that the majority of BBH flares do not occur in the most luminous/massive AGNs (Veronesi et al 2023) hence observations of low mass AGNs, where EM counterparts are expected to be fainter and AGN catalogs are incomplete, are needed. Further aspects that may render the counterpart even fainter include geometrical effects such as the direction of the kick and dust attenuation due to the presence of a dusty torus. The top 5% of BBH events based on their sky localization in O5 match really well the Rubin FoV, as they can be covered with 1-3 Rubin pointings. To conclude, BBH EM counterparts in O5 can be extremely faint, and Rubin will be crucial in a systematic search for those with a handful of exposures and a minimal amount of time.

## Rates and Target of Opportunity activation criteria

**Event rate (triggers/year):** 7 during O5

**Trigger time distribution or constraints:** All during O5.

### Trigger Criteria

- *90% credible interval area* $\Omega$*<20 sq deg,*
- *distance <6 Gpc,*
- *total mass>50* $M_\odot$

We focus on higher mass events since these are more likely to form in a hierarchical environment such as that of AGN disks. We also cut out events at the largest distances to focus on the smallest localization volumes, where the probability of chance coincidence with non-associated AGN flares is smaller. Based on realistic O5 simulations, with the following cuts: 90% credible interval area $\Omega$<20 sq deg, distance <6 Gpc, total mass>50 $M_\odot$, we expect ~22 events per year. Unless in O5 the LVK will release a flag for massive binaries, we will select events based on estimated mass as in Graham et al., 2020. We recommend following up ~⅓ (only those that fall within the Rubin footprint so that we can take advantage of archival Rubin observations to model previous AGN activity and identify with more confidence unusual flaring activity from BBH mergers), ~7 events per year, throughout the duration of O5.

For strongly lensed BBHs, we expect ~1 in 1500-1700 events to be detected with multiple images for LVK sensitivity between design and A+ (Wierda et al., 2021), a range that should include O5 sensitivities. This corresponds to up to 1 event observable by Rubin over the entirety of O5. Given the low probability of such a trigger, we include this one trigger within the number of BBH triggers suggested above and do not recommend extra time for it.

## Time budget and observing strategy

BBH EM counterparts could be AGN flares that can peak over a variety of timescales, 1-~50 days post-merger, depending on the characteristics of the AGN and the binary system. Due to the uncertainty of this case, we should concentrate on significant, very well-localized GW events ($\Omega$ < 20 sq deg) that could be covered with up to 3 exposures, and adopt a log-scaled cadence. We need at least 2 filters to help discern the counterparts from other nuclear transients e.g. nuclear Supernovae (SNe) and Tidal Disruption Events (TDEs). We recommend 30s exposures to keep the depth consistent with archival Rubin data for AGN variability modeling. Ideally, we could have a first pass as soon as possible to eliminate active flares (current models cannot explain a flare that starts before the GW event) and then complement the Rubin cadence in the region so we can both follow up fast flares and identify early any slower flares.

**Area**: Based on our O5 simulations, we consider that 3 events may be covered with only one Rubin pointing, 2 would require 2 pointings, and 2 would require 3 pointings. With 30s exposures for 3 filters, this would take 5 epochs, for 7 events. With 30s exposures for 3 filters, and 5 epochs, for 7 events we recommend a **total of 4.3 hours per year,** including overheads.

For the rare case that a candidate strongly lensed BBH is detected, we will select it based on the detection of multiple GW images as multiple (at least 2) significant events, and specifically on their sky maps overlap statistics (Wong et al., 2021) between multiple skymaps having a false positive rate < 10^-2. The overlapping skymap is what we would follow up on, and it has to satisfy the requirement of $\Omega$ < 20 sq. deg. as for other BBH events, but we do not require it to fall within the Rubin footprint, given the rarity of such an event. We only recommend one trigger of this kind over the entirety of O5, should it occur, it will be considered as part of the total time recommendation above.

**Filters**: If it is not bright time, we recommend *u*- and *g*-band as several BBH in AGN models predict a blue fast component and also we want to discern them from TDEs. TDE studies show that *u*-band rest frame is ideal for identifying these objects. At the typical distances of these events, *u* or *g* should cover the rest frame *u*-band. We suggest the use of *ugi* for events at z<0.4 and *gri* for more distant events as baseline and switch to *riz* in bright time.

**Timing**: The first epoch should be taken as soon as possible to get a baseline - the AGN should not be flaring at that time yet: if the merger occurred within the disk photosphere, it will emerge over a timescale which is around the photon diffusion timescale. This timescale for emission is expected to be highly dependent on the binary properties and the environment, hence we propose to observe on a log-spaced timescale as proposed in the BBH case presented in Andreoni et al., 2022, e.g. days 1-3-8-10-40 after trigger. We will complement ToO observations with nominal cadence LSST observations. Except for the first epoch, if regular survey observations occur in the region of the GW event in one or more of the proposed filters the day before the planned ToO epoch, that ToO epoch can be skipped, hence our time recommendation can be considered conservative. Longer (>40 days) timescales (which may be relevant for the events at the largest distances) will be more likely to be probed by nominal survey cadence, hence we do not propose for those.

## Science case justification and description: Unidentified GW Sources

*Science case lead: Ryan Chornock*

The goal of this trigger is to be sensitive to any type of unidentified GW source class that is not incorporated into the compact binary cases above. Practically speaking, LVK is searching for signals with two different types of searches: template matches for compact object inspirals and burst searches. Potential source classes for the burst searches include (a) Local Group supernovae, (b) neutron star events, (c) binary black holes with masses somewhat higher than the typical ones seen by LIGO (only the merger+ringdown but not the inspiral is in the LIGO frequency band), and (d) the completely unknown. For a very high significance, clearly real event with signal in multiple detectors we will employ either of the Gold or Silver strategies outlined above depending on the sky area. The rates are expected to be low and we recommend up to **16 hours** (the time recommended for the Silver skymap strategy above) for *one* trigger, which is within the uncertainty of our overall time recommendation.

## Total ToO time budget recommended for all the GW/MMA science cases

| **GW case** | **Total time (hrs) - O5** |
|---|---|
| BNS/NS-BH | 240 |
| Lensed BNS | 31 |
| BBH | 11 |
| Unidentified GW | 16 |
| **Grand total** | **298** |

**Table 4:** Total Summary recommendation for GW MMA case during the LVK O5 run

## Emerging Ideas Science Case: Gravitationally lensed Gamma-Ray Bursts and their afterglows

*Science case lead: Graham Smith*

The first discovery of a gravitationally lensed gamma-ray burst (GRB) signal from the distant universe will be a major and very exciting breakthrough, with scientific impact at least comparable with the first discoveries of lensed supernovae (Kelly et al., 2016; Goobar et al., 2017). If the lensed GRB is associated with a binary compact object merger and detection occurs when the LIGO-Virgo-KAGRA (LKV) gravitational wave (GW) detectors are operating, then the scientific impact will rival that outlined in the gravitationally lensed BNS section of this report. This is because the lensed GRB detection may be accompanied by detectable lensed GW signals. Independent of the potential synergy with gravitationally lensed GWs, the first secure discovery of a lensed GRB will open new scientific opportunities in its own right. For example, the sub-second timing accuracy of GRB detectors will enable a significant advance in time delay cosmography as an independent probe of $H_0$ that is independent of the distance ladder. In brief, the timing accuracy will eliminate one of the main error terms in time delay cosmography experiments (Birrer & Treu 2021). The first discovery will also be a novel and unprecedented opportunity to probe the physics of GRBs because the multiple images of the lensed GRB will probe different lines of sight to the central engine and thus different lines of sight through the GRB jet. In addition to these arrival time difference- and deflection angle-related opportunities, gravitational magnification will expand "gravitational telescope" studies of the distant universe to include distant gravitationally magnified GRBs of all durations and variants, and their respective connections with core-collapse supernovae and binary compact object mergers.

Rubin/LSSTCam is uniquely powerful for securing the discovery of a lensed GRB thanks to its unique combination of large collecting area, large field of view, and fast slew speed. Rubin's specific and game-changing contribution will be to detect the faint optical afterglow of a gravitationally lensed GRB and thus localize it to a gravitationally lensed host galaxy in the distant universe (redshift of z>1). The critical role of localization in unlocking the science can be seen in the headline numbers that describe GRB detections to date and the gravitational lensing optical depth. *Fermi* GBM alone has detected >3000 GRBs to date, with typical sky localization uncertainties of up to ~1000 degree$^2$, and of which just ~20% have arcsecond localizations via detection of an afterglow, mostly because of their co-discovery with *Swift* (Connaughton et al., 2015; Goldstein et al., 2020). The theoretically well-defined prediction that roughly one per thousand extragalactic sources of any flavor are gravitationally lensed (e.g. Smith et al., 2023 and references therein) indicates that ~one of the GRB detections to date is gravitationally lensed. However, with no confirmed detections of lensed GRBs in searches to date (e.g. Ahlgren & Larsson 2020; Chen et al., 2022), it is clear that the actual discovery (i.e. recognition) rate is suppressed by the low rate of afterglow detection. Afterglows are challenging to detect within large GRB sky localization uncertainties because afterglows are faint and fade quickly. Indeed, the afterglow detections achieved within hours of GRB detection are dominated by bright sources, AB<~21, and are typically associated with long GRBs that are associated with stellar collapse (Kann et al., 2011). This points to the need to go deep and wide with Rubin ToOs in order to unlock this science.

We envisage Rubin ToO's that aim to detect the lensed optical afterglow to candidate lensed GRBs. Our current estimate is that there will be no more than one candidate and therefore no more than one ToO trigger per year. After taking into account plausible lens magnifications we expect that the regular 30-second WFD depth will be sufficient in each of three approximately log-spaced epochs. Our current thinking on selection criteria is to be inclusive of both short- and long-arrival time differences between the lensed GRB images. The short case would involve selecting individual GRB detections whose temporal and spectral structure presents convincing evidence of comprising two magnified images of a single burst signal. The long case would involve cross-matching new GRB detections with previous GRB detections to identify pairs with matching sky localizations and temporal/spectral structure. Crucially, we would only trigger a ToO for candidates without an afterglow detection from co-discovery with *Swift* and/or a smaller telescope, i.e. reserving Rubin/LSSTCam for the region of parameter space in which it is uniquely powerful. We would also recommend an upper limit of $\Omega <= 1000$ degree$^2$ on any candidate for which we trigger a ToO. Putting all of this together and assuming two filters per epoch (likely *g*- and *i*-bands), we estimate no more than one Rubin ToO trigger per year that totals ~**9 hours**.

In summary, we recommend that the ToO program is flexible enough to accommodate emerging ideas such as that outlined here, as they mature into fully-fledged cases. We envisage completing several projects before putting this forward as a full case, including simulating lensed GRB and afterglow signals, more thorough estimates of the predicted rates, applying our proposed selection criteria to the catalog of GRB detections to date and developing and testing software to identify candidate lensed GRBs in real-time ready for Rubin ToO triggers.

## Science Case: High-Energy Neutrinos

*Science case lead: Robert Stein*

### Science Case justification and description

A flux of high-energy (TeV-PeV) astrophysical neutrinos was discovered by the IceCube Neutrino Observatory in 2013 (The IceCube Collaboration, 2013), providing evidence for the existence of extreme cosmic ray accelerators, whose origin has not been clearly identified yet. Some neutrinos arise from our own Milky Way (The IceCube Collaboration, 2023), but most neutrinos have an extragalactic source. The nearby active galaxy NGC 1068 is the sole high-significance steady neutrino source in the sky (The IceCube Collaboration, 2022). Many time-varying sources of high-energy neutrinos have been proposed, as illustrated in Figure 4.

Since 2016, IceCube automatically reports the detection and localization of individual high-energy neutrinos in near realtime (Aartsen et al., 2017; Abbasi et al., 2023), enabling fast multi-wavelength follow-up campaigns to discover these time-varying neutrino sources. The IceCube neutrino alert program has been very successful, leading to the identification of the flaring blazar TXS 0506+056 as the first probable TeV neutrino source (The IceCube Collaboration, 2018), and more recently optical follow-up has identified both TDEs (Stein et al., 2021, Reusch et al., 2022) and a rapidly-evolving interacting supernova (Stein et al., 2023) as probable sources of high-energy neutrinos. This represents the first evidence confirming the long-held predictions that these populations are cosmic particle accelerators. However, most of the cosmic neutrino flux remains unexplained.

The identification and characterization of populations responsible for the astrophysical neutrino flux would be a milestone in multi-messenger astronomy. Neutrinos offer an entirely new window into optical transients, carrying unique information about the acceleration mechanisms and interactions that occur within these cosmic accelerators. They probe high-energy processes which are typically completely invisible in photons. Neutrinos are thus the best way for us to conclusively identify and study cosmic accelerators. Optical follow-up programs have a proven record of success, but current instruments are only capable of detecting the very brightest sources. Follow-up with Rubin would be truly transformational, providing a comprehensive view of the multi-messenger optical sky.

### Justification for the use of Rubin

Neutrino telescopes are capable of detecting individual high-energy neutrinos from throughout the entire observable universe, rather than being heavily biased towards the nearby universe in the same way that electromagnetic surveys are. This presents a particular and unique challenge for neutrino follow-up, many of the neutrino alerts will originate from the high-z universe.

As a consequence, follow-up programs with existing programs like ASAS-SN/ZTF are blind to the majority of high-energy sources (see Figure 5). Existing programs are only sensitive to nearby or bright objects, and so for any individual alert, the probability for these facilities to detect a counterpart is low. Indeed, the handful of time-varying neutrino sources identified to date share a common property: all were very luminous at optical wavelengths.

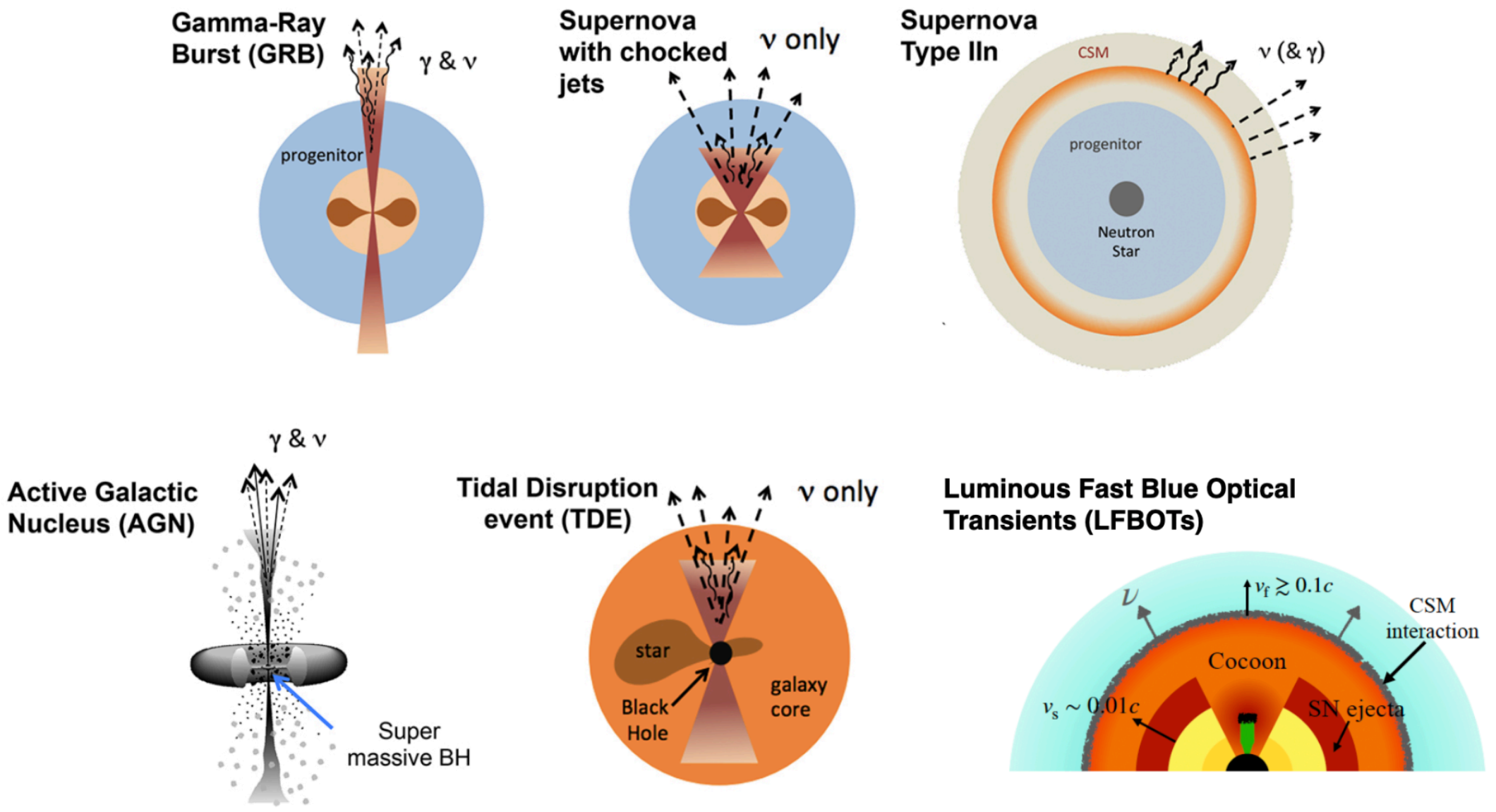

**Figure 4**: Proposed time-varying neutrino source classes, adapted from Bartos (2018) and Guarini et al (2022). All six populations would have time-varying optical emissions, and our proposed Rubin ToO program would be sensitive to all six populations.

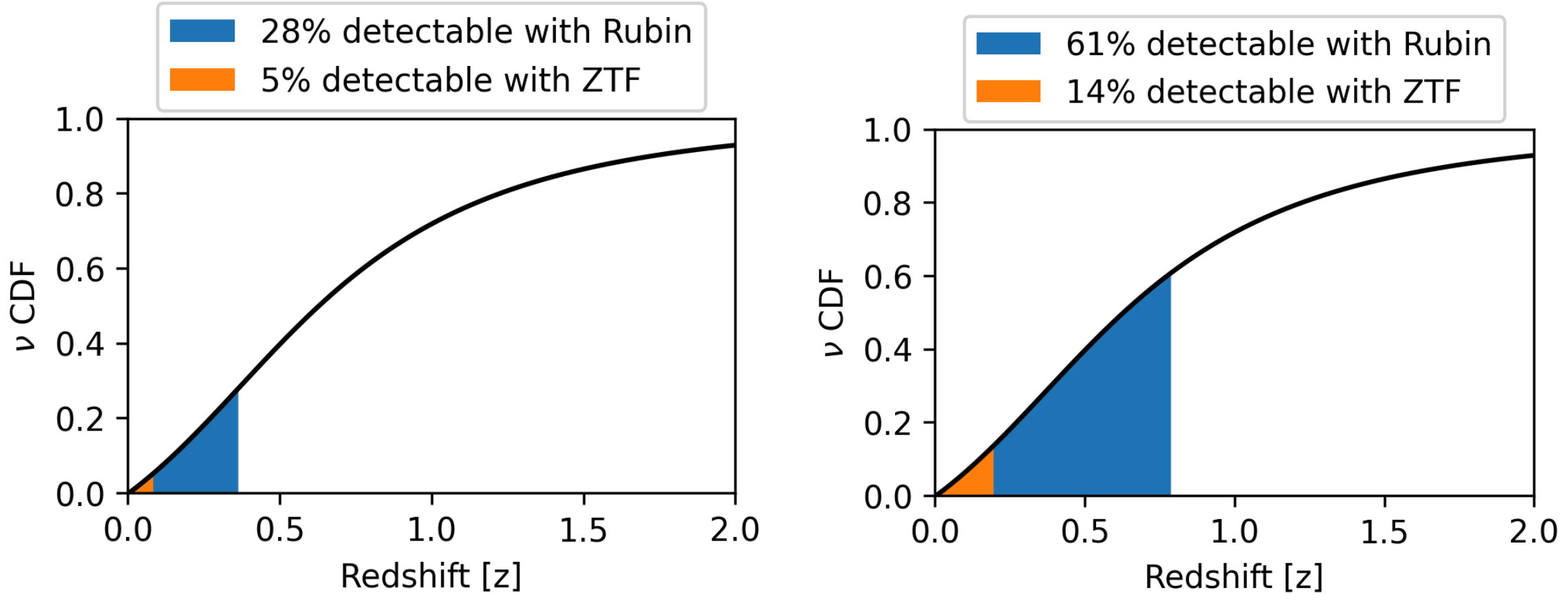

**Figure 5:** The neutrino cumulative distribution for neutrino sources following the star formation rate (black). The median redshift for a neutrino tracing the SFR is z~0.6. **Left**: For a source with M=-17.5, ZTF would have a 5% chance of detecting the counterpart, while Rubin would have a 28% chance. **Right**: For a brighter source with M=-19.5, ZTF would have a 14% chance of detecting a counterpart, while Rubin would have a 61% probability of detecting the counterpart. In both cases, Rubin's chance of detecting each counterpart is more than four times higher.

An ideal discovery tool for electromagnetic counterparts to high-energy neutrinos would combine four properties:

1. Large field of view to cover neutrino contours, which have a median localization of ~5 sq deg.
2. Depth of m>24.5 mag to reach fainter transients at z>0.5

3. Deep high-cadence survey baseline (m>24.0 every ~3 days) to reveal both historical and recent detections, as well as to provide constraining upper limits.
4. Multi-band color information for classification

In combination, this would provide a decent chance to detect the counterpart to any individual neutrino, while also allowing unrelated transients to be efficiently filtered out. Currently operating neutrino follow-up programs include ASAS-SN, DECam, Kiso, Pan-STARRS1, MITSuME, Subaru, and ZTF (The IceCube Collaboration, 2018; Kankare et al., 2019; Morgan et al., 2019; Necker et al., 2022; Stein et al., 2023). However, none of these surveys meet all four criteria.

Fortunately, *Rubin is an ideal neutrino counterpart discovery machine,* uniquely combining all required characteristics. Only Rubin can be used to discover the many distant neutrino sources. For unremarkable optical magnitudes of M=-17.5 or M=-19.5, the chance of detecting a neutrino source with Rubin is more than four times higher than e.g. ZTF (see Figure 5). Rubin will thus be the first instrument capable of identifying neutrino sources amongst these populations.

## Rates and Target of Opportunity activation criteria

**Event rate (triggers/year)**

Min: 2 (expect 3-4 on average)

Max: ~12 (estimated based on future observatories coming online)

**Trigger time distribution or constraints:** IceCube is running already, with 99.5% uptime, and is expected to remain online through the Rubin LSST survey. With IceCube, we expect ~4 triggers per year that are accessible to Rubin and meet the trigger criteria. New facilities under construction (e.g. KM3NeT) have a similar size to IceCube, so the base neutrino alert rate should be similar to IceCube. However, these new facilities will be based in water rather than ice, and are expected to have a better average localization. Therefore, we expect a larger fraction of these new alerts to pass our trigger criteria. In contrast to IceCube, these new instruments will detect neutrinos preferentially in the Southern hemisphere, meaning most are likely to be accessible to Rubin.

Neutrinos should be triggered which are well-observable with Rubin, can be covered with a single pointing, and have a high probability of being astrophysical. We define our criteria based on IceCube neutrino alerts, but triggers from future neutrino telescopes such as KM3NeT or Baikal-GVD should be subject to the same simple algorithm:

### Trigger Criteria

- *Latency to first Rubin exposure: <24 hours to observation*
- *P astro: >50%*
- *Localization: >70% of reported contour (including sys unc) covered by a single Rubin pointing*
- *Galactic Latitude: >10 degrees*
- *Neutrino in Rubin Footprint*

- *No obvious neutrino data quality issues (neutrino alerts are occasionally retracted, and ToOs should not be performed if an alert is retracted).*

IceCube is the only cubic kilometer neutrino telescope currently operating and producing public alerts, so we explain the procedure for this instrument as of 2024. The workflow may need to be slightly tweaked for other instruments that come online, or if the IceCube reporting procedure changes.

**We can trigger based on the simple algorithm above. However, the information required to trigger is not yet published in an automated machine-readable way.**

At present, IceCube employs the following reporting procedure:

1. The detection of high-energy neutrinos is reported automatically by IceCube via machine-readable GCN Notice, with minimal (60-120s) latency. This includes the estimated astrophysical probability ("signalness"), which does not get updated. It also includes an initial localization, accounting only for statistical uncertainty but not systematic uncertainty.
2. Further reconstruction algorithms are employed by IceCube to provide a more accurate localization accounting for systematic uncertainty. This takes approximately 2 hours but varies based on event properties.
3. In the meantime, designated IceCube collaboration members vet the quality of the events.
4. IceCube summarises the properties of the event using a plain-text "GCN Circular". This includes an updated localization. The final localization of a neutrino is typically irregular rather than a circular Gaussian, and IceCube reports a bounding rectangle fully containing the 90% contour. IceCube also reports additional contextual information in an ad-hoc manner, including possible additional indications of a background nature that are not captured by the "signalness" parameter. The coordinates of the bounding rectangle are also included only via human-typed text, so the exact format is not always uniform between different events. As a consequence, this information cannot automatically be parsed via a machine.
5. IceCube does release an updated machine-readable GCN notice revision, reflecting the new localization. However, the GCN notice uses a circular approximation of the position, and therefore these GCN Notices have degraded resolution for assessing coverage.

We envision the following workflow for Rubin:

1. Initial neutrino alerts are received from IceCube or other observatory. This provides an early warning for a potential trigger.
2. Rubin should await the final neutrino localization with systematic uncertainty.
3. Based on the updated alert, code executed by Rubin will calculate the final area for the neutrino, and whether that area can be substantially covered with a single Rubin pointing.
4. Using that information, and accounting for the observability of a particular trigger: an algorithmic decision can be made on whether a given trigger is "Go" or "No Go" (with oversight from the appropriate committee).

5. If “Go”, the optimal center for the single ToO field should be chosen to maximize enclosed probability.

**We strongly encourage the IceCube collaboration to implement the automated release of fully-machine-readable neutrino alerts, including localizations with systematic uncertainties and a simple integer data quality flag, to enable a neutrino trigger decision algorithm for Rubin. The IceCube collaboration has expressed a willingness to implement this. We also strongly encourage other neutrino observatories to release public alerts in a similar format.**

We suggest that the overall trigger threshold be reevaluated if/when new observatories approach design sensitivity.

## Time budget and observing strategy

Quantitative goals for the follow-up:

- **Depth**: Nominal survey depth in u,r,z as well as three deeper 120s exposures in g
- **Time**: Aim to trigger first observation ASAP, within 24 hours of event time
- **Area**: One single Rubin pointing, centered to enclose or maximize the coverage of the reported localization area
- **Filters**: Filters *grz* in predefined sequence, plus *u*-band as soon as it is mounted

**Search Area (sq deg)**

Min: 0.5

Max: ~6 (Always use a single field, trigger if a Rubin ToO would get >70% coverage of contour.)

Typical: 3

**Max delay from trigger (hours):** As illustrated in Figure 4, there are several possible source classes that our observations must be sensitive to. Early observations are essential to detect GRB afterglows and jetted TDEs and to constrain the explosion of a supernova with a choked jet. They are also essential to resolve rapid AGN flares, which can fade on timescales of hours. LFBOTs can fade within a couple of days, while non-jetted TDEs and regular SN are not substantially affected by a delay in ToO observations. To summarize, there is no maximum delay that restricts the proposed follow-up, but our sensitivity to different neutrino source classes decreases.

### Observing Strategy Details

We model our recommendation after the strategy employed by ZTF, which combines a multi-band baseline survey with deeper ToO observations. Rubin will have vastly greater depth (see Figure 5), and much broader wavelength coverage, but the baseline optimization is similar:

- Day 0 to obtain a deep baseline and identify currently bright sources
- Day 1 to detect sources that are rapidly rising or fading (GRBs, LFBOTs, Jetted TDEs) in particular

- Day 7 to identify sources over intermediate timescales. This ensures all slower-evolving candidates can be photometrically classified.
- Day X *u*-band as soon as available to identify hot transients e.g TDEs/LFBOTs

Each day would combine multiple filters, to enable promising candidates to be identified via color, color evolution, and rapid light curve evolution. Beyond this, we would rely as much as possible on the baseline survey to monitor slower-evolving transients.

The aim of the observation strategy is to enable all major populations in Figure 4 to be detected and to further enable photometric classification so that members of each population can be identified. Because these populations vary on different timescales, we require our observations to be distributed over multiple nights.

**Day 0:**

We suggest beginning observations ASAP after neutrino localization, with a single deep observation in *g*-band. (120s)

We also recommend *r*-band (30s) separated at least 15 minutes from *g*-band, and *z*-band observations (30s) at some point in the night. We do not have strict conditions for the timing of the *z*-band observations. *r*-band is important to provide a second detection for most transients, *z*-band to identify extremely red sources (e.g. GRB afterglows).

**Day 1:**

We would again observe in *g*-band (120s), and *r*-band (30s). This will resolve rapidly evolving transients, sensitive both to fading/rising and color change.

**Day ~7 (with some flexibility to minimize survey disruption):**

We would observe one final time in *g* (120s), *r* (30s), and *z* (30s) to detect a wide variety of transients that may have evolved since day 0 in color or luminosity.

**Day X**

We additionally recommend one *u*-band observation (30s) for color information, either on day 0 or whenever the *u*-band filter is next mounted. *u*-band is a powerful method to identify TDEs/LFBOTs, and we expect a typical median latency of ~1 week.

In total, assuming only IceCube, we expect ~4 triggers per year, each covered by a single pointing. For a given neutrino, we recommend multiple visits on night 0, 1, 7, and X. In total, we recommend a total exposure time (6x30s) + (3x120s) **= 540s per neutrino, corresponding to 36 minutes per year for 4 triggers.** If future detectors increase the alert rate to e.g. 12 per year (similar volume = similar alert date), we would recommend **~100 minutes per year plus overhead.**[13]

**Fraction of triggers that need day 1 ToO:** 100%

**Fraction of triggers that need day 7 ToO:** 100%

[13] Because the strategy is flexible and time gaps between images in different filters, the overhead is difficult to estimate. Optimistically, the filter changes fall naturally in the non-ToO night observing plans, pessimistically, assuming 120s per filter change, they may ~double the requested time.

**Search area per day after day 0:** A single Rubin pointing on all days

### Scientific loss if day 1+ observations are not acquired

There will be many sources detected in Rubin observations, and it won't be feasible to perform follow-up for each. Day 1 observations are essential to identify rapidly-evolving sources, and Day 7 observations are needed to perform photometric classification. Only with the full sequence can promising candidates be feasibly selected for follow-up.

## Special Observing Requests and Data Processing Requirements

We would recommend deeper observations in a single filter (*g*-band), in order to better constrain the rise/fade of transients detected with the nominal survey cadence. In particular, neutrino models invoking jets at the time of supernova explosion can be excluded with deep detections of a rising supernova shortly after neutrino detection. We have adopted 120s as a nominal exposure time for these three deep observations to achieve this aim, and estimate that these observations would have a depth of 25.8 mag AB. However, we would adjust this based on the results of the commissioning performance.

This ToO strategy does not require any specialized data processing.

We recommend sufficiently high priority to observe even during an ongoing GW trigger. The impact on GW science would be minimal.

## Downscoping direction suggestions

The neutrino program would only be triggered on events for which a single Rubin pointing was sufficient, and it is therefore not possible to descope through fewer fields. Similarly, the cadence recommendation is the minimum required to be sensitive to all target neutrino source populations, and we would strongly discourage any descoped cadence for an individual trigger.

One way to descope the program would be to restrict the overall number of neutrino triggers. However, we caution that the origin of high-energy neutrinos can only be established through a statistical analysis, and this requires a large sample of ToO triggers.

The most reasonable way to descope the program would be to support "reweighting" of the standard Rubin survey cadence to achieve the desired neutrino cadence with minimal disruption to the overall survey. We would encourage the Rubin Observatory to support less disruptive ToO requests (e.g. allowing a single filter observation to be conducted within a window of +/- 1/2 days, at a time when the survey is already observing in said filter).

## Total ToO time budget recommended for the neutrino science case

| | |
|---|---|
| **Nu case** | **6-18 hours** |

## Science Case: Galactic Supernova

*Science case leads: Robert Stein, Claire-Alice Hébert, John Banovetz*

### Science Case justification and description

No Galactic core-collapse supernova (CCSN) has been discovered in real-time in nearly 1000 years, and there has only been one nearby SN in recent decades. Estimated Galactic SN rates are 1-2 per century (Rozwadowska et al., 2021), so there is a ~15% chance of one exploding during the 10-year survey. This presents a rare and exceptional opportunity to observe such an event from a front-row seat.

In the era of multi-messenger astronomy, we will be able to schedule observations of a galactic SN before the first electromagnetic signal even reaches us. Neutrinos and potentially gravitational waves will contain rich and complementary information about the astrophysics of SN explosions. Specifically, as shown by SN1987A, a CCSN's neutrino signal is anticipated to arrive tens of minutes to several tens of hours ahead of the shock breakout (SBO). SBO occurs when an explosion reaches the star's surface, with the escaping photons resulting in electromagnetic brightening in optical and infrared bands (Hirata et al., 1987; Kistler et al., 2013). The initial neutrino burst will provide an alert of the event, approximately localize the SN, and also set a reference from which to measure the time delay to SBO; this time delay depends on the progenitor system, as shown in Figure 6.

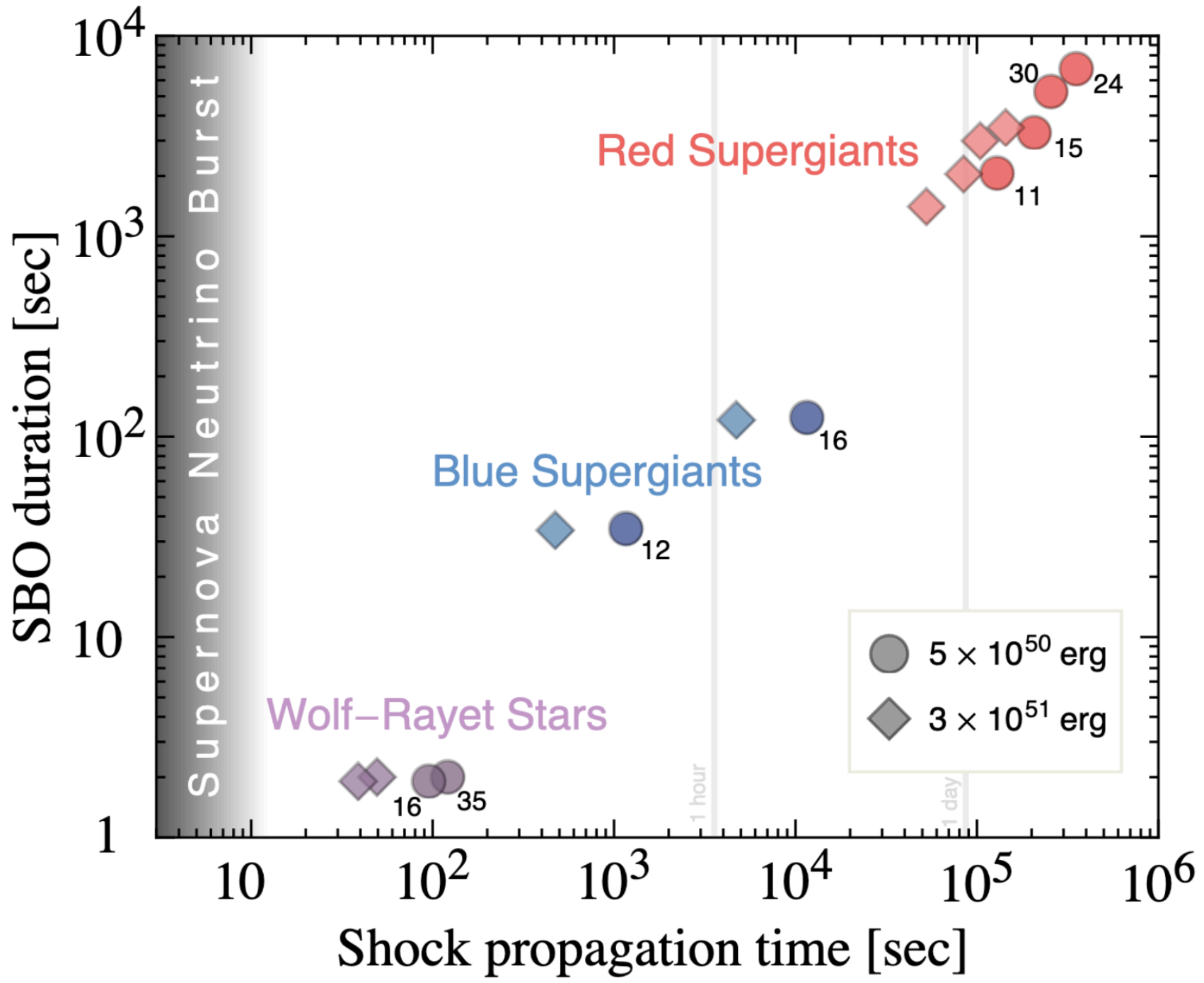

**Figure 6:** The horizontal axis shows shock propagation time, i.e., the delay between a SN neutrino burst and the start of SBO, calculated for a variety of initial progenitor masses and shock energies. More details in Kistler et al., (2013). For a galactic SN, the time delay from neutrino burst to SBO could be from as little as two minutes to as long as two days.

Most (extragalactic) SN are discovered near peak brightness, so we have little information about the very earliest parts of the light curve, i.e. SBO and the ramp-up to peak brightness. Prompt and comprehensive follow-up of a galactic supernova will reveal unprecedented insights into the explosion mechanism of the SN and information about the surrounding material (Bersten et al., 2018; Taddia et al., 2015), as well as potentially enable spatially resolved imaging of early phases of the SN explosion. To take full advantage of such a once-in-a-lifetime event, it is crucial that we identify the optical counterpart that can appear within minutes of the neutrino alert.

## Justification for the use of Rubin

The combination of a large field of view and great depth make Rubin an ideal instrument for locating a galactic SN (Christopher, Walter., 2019).

Neutrino detectors will provide an alert prior to any electromagnetic (EM) signal through the Supernova Early Warning System (SNEWS) or direct communication with detectors capable of providing useful pointing information (currently only Super-Kamiokande; Hyper-Kamiokande and DUNE are expected to turn on during LSST) (Al Kharusi et al., 2021). For a typical core-collapse SN at 10kpc from Earth, for example, we expect a pointing accuracy of ~25 square degrees from Super-K, with pointing improvements scheduled. Rubin Observatory's wide field of view can cover the Super-K localization region within just a few pointings, giving us the best chance of catching SBO and enabling rapid identification that will be invaluable to the larger astronomical community, allowing a broader set of dedicated and focused telescopes to observe this event at many wavelengths.

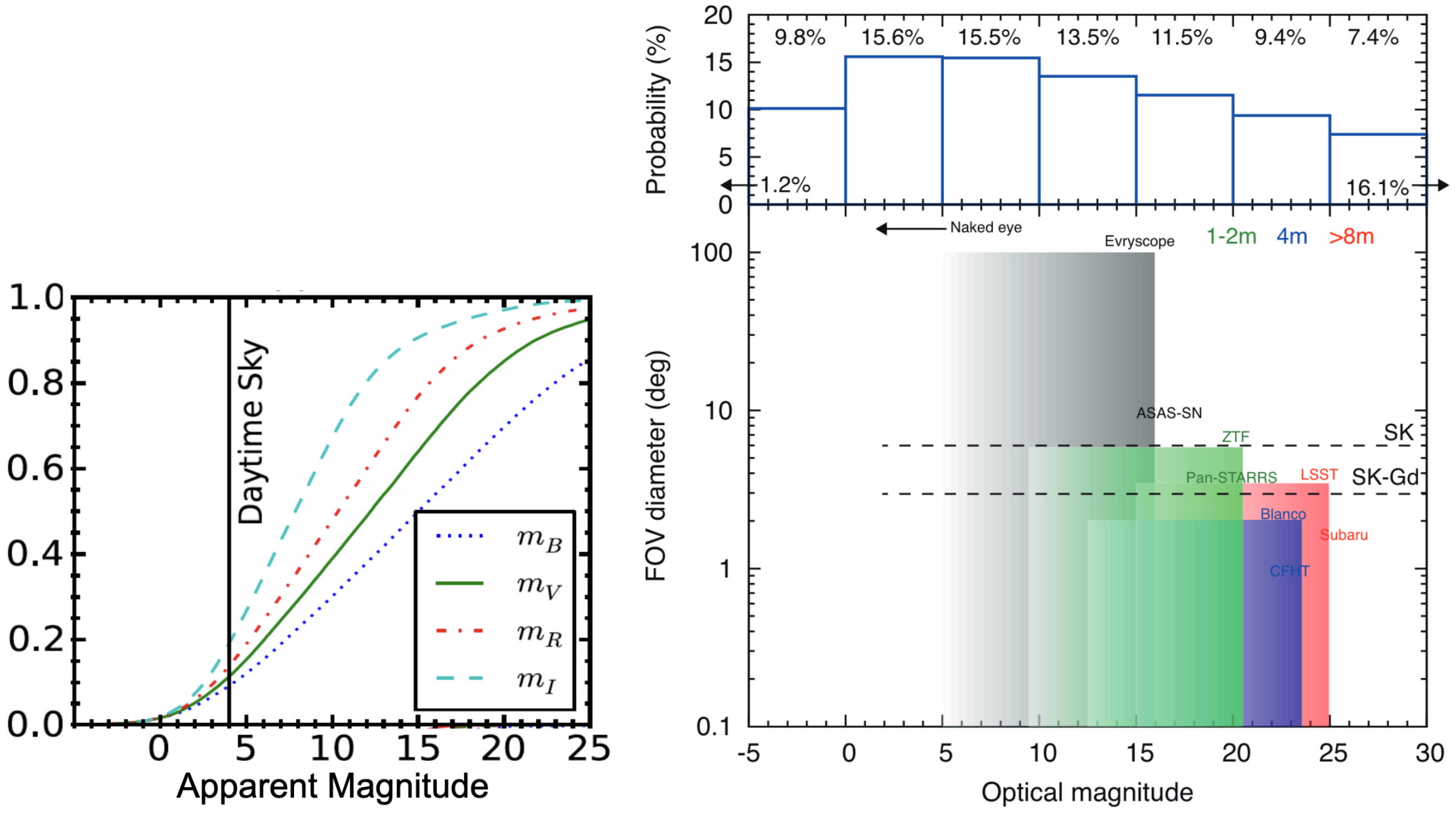

**Figure 7**: **Left**: Cumulative SBO magnitude probability distribution in various optical bands, from Adams et al., 2013. The bluer emission from SBO may only be discoverable with Rubin. **Right**: Detection prospects for galactic core-collapse SN, from Nakamura et al., 2016. The top panel shows the peak SN magnitude probability distribution in the optical band; the bottom panel compares the detection capabilities of various telescopes, with Rubin Observatory in red. Horizontal dashed lines represent the approximate pointing resolution of Super-K; the lower "SK-Gd" line is current since 2020.

The SN and SBO magnitude probabilities, as well as the Rubin detection capabilities, are summarized in Figure 7. The SBO signal is likely to be very bright; it will almost certainly peak above 25 mag in the *i*-band. This is the first signal we aim to detect.

If we miss SBO, we still aim to identify the counterpart before peak brightness. While one might naively expect a nearby SN to be bright, it is likely to suffer from very heavy extinction depending on where it is located in the Milky Way. There is a 25% chance that the SN will be visible to the naked eye at peak, but a 15% chance of it peaking below 21 mag in the optical. Importantly, we want to detect the SN many magnitudes below peak brightness in order to learn more about the progenitor from the early light curve; this detection can only be achieved with the depths provided by Rubin.

## Rates and Target of Opportunity activation criteria

**Event rate (triggers/year)**

Min: 0

Max: ~0.02; over the course of the full survey there is a ~15% chance of receiving one alert, and a 1% chance of a second.

**We should trigger Rubin ToO Observations for the next galactic supernova alert.**

**Trigger format:** The particulars may vary, but we expect both a SNEWS alert (false alarm rate < 1 per century) and a localization from Super-K. The exact format of the notification will depend on the source of the trigger. The SuperK collaboration meeting plans to release a GCN Circular, as detailed in [6]. SNEWS 1.0 is currently operating, and the updated SNEWS 2.0 is expected to be operating in time for Rubin. We anticipate that SCIMMA will be capable of providing the localization in machine-readable format, but emphasize that Rubin should adapt as supernova alert infrastructure evolves.

## Time budget and observing strategy

The delay between neutrino burst and shock breakout can be up to 1-2 days. Rubin should remain triggered for the duration of that window, as long as no counterpart has been found.

Quantitative goals for the follow-up:

- **Depth**: We aim to detect at least one of SBO (bright) or the early parts of the rising light curve (faint); alternating short and long exposures while tiling could address both searches. As seen in Figure 7, there is a range of possible magnitudes to target; the SN will most likely be below 15th magnitude, however could be as deep as 25th mag due to extinction. The exposure depths should be dynamic once the search is underway, as specific information from the Super-K localization may inform this choice.
- **Time Frame relative to trigger**: As quickly as the observatory can switch observing modes, as shock breakout could be minutes to days and non-detections give information about the progenitor system
- **Area**: Max area likely about 100 square degrees, typical area of 25 square degrees
- **Filters**: Observe with *i* band to minimize the effect of extinction.

**Search Area (sq deg)**

Min: 9

Max: 100

Typical: 25

The best and most likely case is that Super-K (present) or Hyper-K / DUNE (future) are online when the SN explodes, as these detectors will provide the best localization for SN follow-up. In the unlikely event that Super-K is down, the neutrino pointing would be constrained from triangulation of neutrino arrival times at other detectors, leading to a much larger tiling region to search.

**Max delay from trigger (hours)**

Ideally, we would trigger within minutes of getting the alert. A lot of information about the progenitor system can be discovered through the early light curve. We would also like to observe shock breakout, which can occur minutes to days after the neutrino trigger (see Figure 6 above). If the shock breakout is missed, each hour of additional delay means missing out on crucial details of the progenitor system and the explosion mechanism.

**Trigger time distribution or constraints**

None

### Observing Strategy Details

This science case may require observations up to 48 hours out from the trigger, since the electromagnetic counterpart of the explosion may take two days to appear.

**Prior to detection:**

Continuously tile high-probability area in *i* band *until the counterpart is found*.

Alternate short *i*-band exposures (exact exposure time TBD pending commissioning results) with longer 15-second exposures; short exposures would aim to catch shock breakout and “de-crowd” the field, while longer exposures would aim to catch the rising light curve in case we missed SBO.

**After detection:**

Take 2-3 more exposures in the *i*-band to remove any false positives and confirm the detection, and continue observing in multiple filters while the detection is reported to the community. Coordinate closely with other observatories, especially those with instruments capable of *u*-band or bluer measurements such as DECam. Rubin should definitely continue imaging in *i*-band until more observatories are on-sky observing the SN. After that point, Rubin no longer has an obligation to observe based on its unique capabilities.

However, there could be significant scientific merit in continuing observations in multiple bands for at least the rest of the night to take full advantage of this unprecedented event. Having a high-cadence coverage of the rising light curve without gaps will be invaluable, as will broad wavelength coverage. If these observations cannot be met by the community (for example due to weather, instrument maintenance, a lack of complementary filters or insufficient sensitivity) then continuing observing with Rubin will have considerable value. This will be especially true if the galactic SN is heavily extincted, in which case only large telescopes like Rubin will be capable of detecting the source. The appropriate strategy will depend heavily on the properties of the event itself. We strongly advise that the SCOC/ToO scientists should determine in real-time whether additional Rubin observations are worthwhile, and what their scope should be.

Late-time follow-up of the supernova post-peak should be considered at the discretion of SCOC, based on the properties of the individual event itself and the uniqueness of Rubin's capabilities. However, these decisions will not be required for several weeks after the supernova explosion, and we suggest that specific proposals are formulated at that time.

**Fraction of triggers that need day 1 ToO**: depends on the event

**Fraction of triggers that need day 2 ToO**: depends on the event

**Search area per day after day 0**: If no counterpart yet -> 100 sq deg (exact number depends on the event), otherwise none

Some tiling may be required depending on the distance at which the SN occurs, which sets the Super-K pointing resolution, but we should be able to cover the area within 10 visits (if we only have pointing information from triangulation, the number of visits needed to tile the localization region will be much higher). Depending on the SN type, the shock breakout could take up to two days to occur and may require tiling the area for 1-2 days after trigger.

Scientific loss if day 1+ observations are not acquired:

If there is a small delay between the neutrino trigger and shock breakout (<1 day), we could miss the rise of the SN as well as not be able to improve on the neutrino localization.

## Special Observing Requests and Data Processing Requirements

Depending on the alert, we may need lower exposure times as there is a roughly 50% chance of a galactic SN being brighter than 15 mag (roughly the saturation limit for Rubin). This exposure time should be a dynamic depth; because of extinction, it would depend on both the distance of the object as well as its location in the Milky Way.

This science case may require one-second exposures depending on the location and the brightness of the SN. Even at one-second exposures, we would be sensitive to depths of up to 20-22 mags and the majority of SNe (see Figure 7). If the area is small enough, dithering would also be required to be able to avoid chip gaps.

The requirements for data processing are currently under investigation. Some questions we need to answer are:

1) Whether existing Difference Image Analysis (DIA) algorithms can succeed in extremely crowded fields near the galactic plane/center, where a galactic SN is most likely to be found, or if we need to develop other methods. Work is currently underway to test this.
2) Whether we will be relying on existing galactic templates or generating them on the fly as we localize the SN.
3) If we rely on templates, what to do if we receive an alert in the early months of the survey before they have been adequately constructed

## Downscoping direction suggestions

There are two primary components to galactic supernova follow-up: counterpart discovery and then follow-up observations of a known source. While the case for a Rubin discovery of a counterpart is strong, the need for continuing observations after the counterpart has been found and other instruments, e.g., DECam, have begun taking data is less pressing. We suggest that the scope of any additional observations is left to the discretion of the SCOC based on the particulars of the event.

However, there is likely to be no more than one trigger in the lifetime of Rubin, and there are consequently no obvious descope options in terms of trigger criteria. We strongly recommend that any galactic supernova is triggered on, regardless of neutrino localization. Once localized, continuing observations through 1-2 nights will give consistent light curves and that will be of immense value to the astrophysics community.

# Science Case: Small PHA Potential Impactor

*Science case leads: Tim Lister, Sarah Greenstreet*

## Science Case justification and description

The near-Earth object (NEO) small body population is vital for understanding Solar System formation and evolution, linking to the formation of exoplanetary systems, and for planetary defense. For the latter, the subset of NEOs with diameters $d > 140$ m and orbits that come within 0.05 au of Earth's orbit are traditionally referred to as Potentially Hazardous Asteroids (PHAs). There is good evidence, however, that the much more numerous smaller objects down to $d$~10 m pose a localized impact risk, from events such as the 500 kiloton-equivalent airburst produced by the $d$~17 m object that exploded over Chelyabinsk in 2013 Feb. These $d$~10-50 m objects are larger than the more frequent meteorite-dropping events commonly picked up by smaller-aperture facilities, which are not a priority for Rubin ToOs; they are also smaller than the traditional definition of PHAs, however, for this report, we refer to these objects as "small PHAs" to emphasize their potentially hazardous nature despite their smaller sizes.

Given that sky-plane uncertainties and rate of sky motions, which increase trailing losses, can increase very rapidly (Milani, 1999) within a few days of close-passing small PHA discovery, there is a need for Rubin ToOs on a limited subset of objects discovered that pose a potential threat to the Earth and for which follow-up would be impossible on other facilities (see Rubin-use justification below). A small fraction of these objects could be potential impactors coming from the sunward direction discovered by Rubin's near-sun twilight microsurvey; these sunward-incoming small PHAs could require rapid-response Rubin ToOs before the discovery twilight period ends in order to confirm or rule out the possibility of impact before the object is no longer visible.

## Justification for the use of Rubin

The Solar System ToO science case is unusual, perhaps unique, among those considered in that Rubin itself is extremely likely to be the prime trigger source for Rubin ToOs. This is due to the combination of a larger mirror diameter and larger field of view (4x and 7x the next-largest NEO survey telescope, respectively), location in the opposite hemisphere to the majority of other search assets, and near-Sun survey capabilities that give Rubin an unprecedented edge in the discovery space. Similarly, these same attributes make Rubin the only or most time-effective method of follow-up for small PHAs that exceed the capabilities of the 4-m Blanco+DECam or the 3.6-m CFHT+MegaCam (for objects discovered by Rubin that rapidly move into the Northern Hemisphere). We have been careful to write our trigger conditions below to exclude objects that are likely within the capabilities of these other two possible follow-up facilities (assuming suitable ToO observing modes and rapid pipeline processing facilities are available for Solar System ToO science cases on these non-Rubin facilities).

Rubin's unprecedented depth leads to discoveries of close-passing/potentially impacting small PHAs at much greater distances from Earth than existing NEO survey facilities can accomplish (see Figure 8 and Figure 9 caption). Typically, close-passing/potentially impacting small PHAs are plagued by high rates of motion and increasing time since last reported detection, which both correspond to an increasing likelihood of newly discovered/short-arc object loss through a rapidly increasing ephemeris uncertainty and thus on-sky search area combined with increasing trailing losses over that search area that can exceed all possible capabilities at other facilities without rapid response observations (see Figure 9). For objects with a non-negligible impact probability, this can obviously be catastrophic without orbit-improving follow-up astrometry to remove the impact risk. Rubin's ability to detect these small PHAs at greater distances from Earth and reduce or eliminate their impact probabilities through rapid follow-up is thus required before the above issues become a problem, removing the ability of other facilities to perform the necessary impact-probability-reducing follow-up.

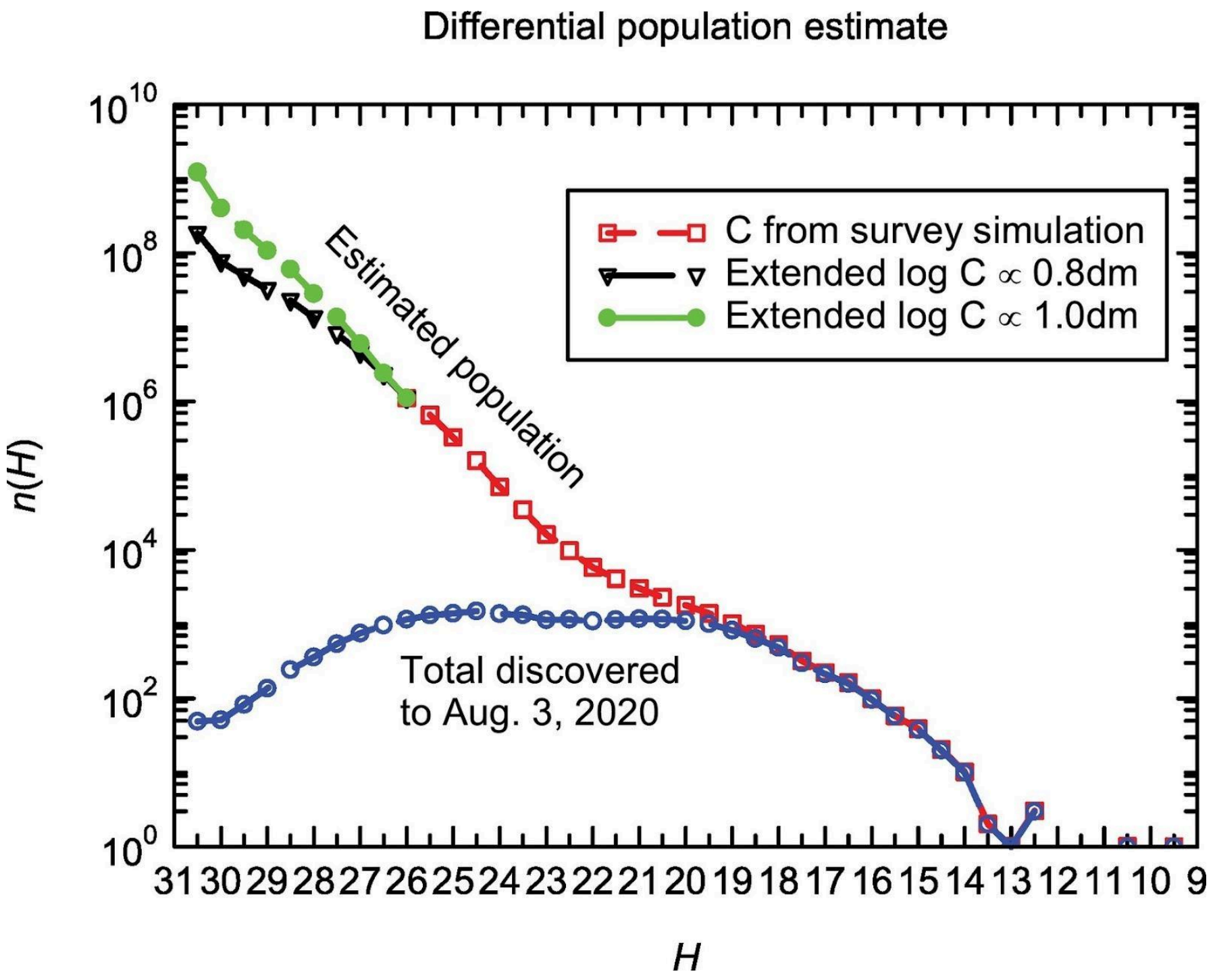

**Figure 8** (from Harris and Chodas, 2021): The differential population estimate of NEOs as a function of absolute magnitude *H* follows a steep power law, for which the normalization at the small-size end (26≲H≲31) has considerable uncertainty but will be substantially improved with Rubin and NEO Surveyor (Mainzer et al., 2023). Rubin's etendue is 28x greater than the next-largest NEO survey facility, which will make Rubin the premier facility for discovering "small PHAs".

A study performed by the Asteroid Terrestrial-impact Last Alert System (ATLAS; a system of 4 x 0.5 m NEO survey telescopes) presented in Tonry et al., (2018) shows that *H*~25.4 (d~30 m) objects are visible to ATLAS at geocentric distances of up to 0.7 au (their Fig 8). Our preliminary analysis with modified metrics and extrapolated NEO population models (since the Rubin PHA model population based on Granvik et al., (2018) has no objects with *H*>25) suggests that Rubin will detect (5-sigma) *H*~28 objects at similar geocentric distances, ranging from 0.04-0.20 au [Lynne Jones, pers. comm]. (More in-depth modeling and "as-operating" measured values of Rubin's performance would be necessary to turn this into a true probabilistic detection volume perpendicular to and along Earth's orbit, e.g the "candle flame volume" of Tonry et al., 2018, Fig 8.) Based on a query of the JPL Small-Body Database API[14], 30 *H*<28 confirmed NEOs passed within 1 lunar distance of the Earth from 2023 March - 2024 March.

[14] https://ssd-api.jpl.nasa.gov/doc/cad.html

Generating ephemerides for Rubin for these 30 $H$<28 objects, we find an extremely broad range of magnitudes ($V$~20–31) at t-10 days before close-approach corresponding to the approximate center of the Rubin detection distance range above (0.04-0.20 au). (We have neglected the $V$-filter transformation here but [Vereš et al., 2015](), quotes $V$-$g$=-0.28, $V$-$r$=0.23, $V$-$i$=0.39, $V$-$z$=0.37 for a mean S+C NEO taxonomy for the very similar PanSTARRS filter system). This magnitude range obviously corresponds to very easy for Rubin (and other follow-up facilities) to impossible even with much deeper than standard Rubin exposure times. As the newly discovered NEO approaches Earth, it obviously increases in brightness but also in the on-sky rate of motion. Based on the ephemerides for the 30 $H$<28 objects described above, the range of motion at t-10 days before closest approach is 0.1 – 0.75 deg/day, which results in trailing losses of 0.006 – 0.25 mag (assuming $r$-band FWHM=0.8" and 30s exposures). At the time of closest approach, on-sky motion ranges from 10.9 – 650.7 deg/day with a median of 112.7 deg/day, well above the security-mandated maximum of 10 deg/day, which results in 2.5 mags of trailing loss.

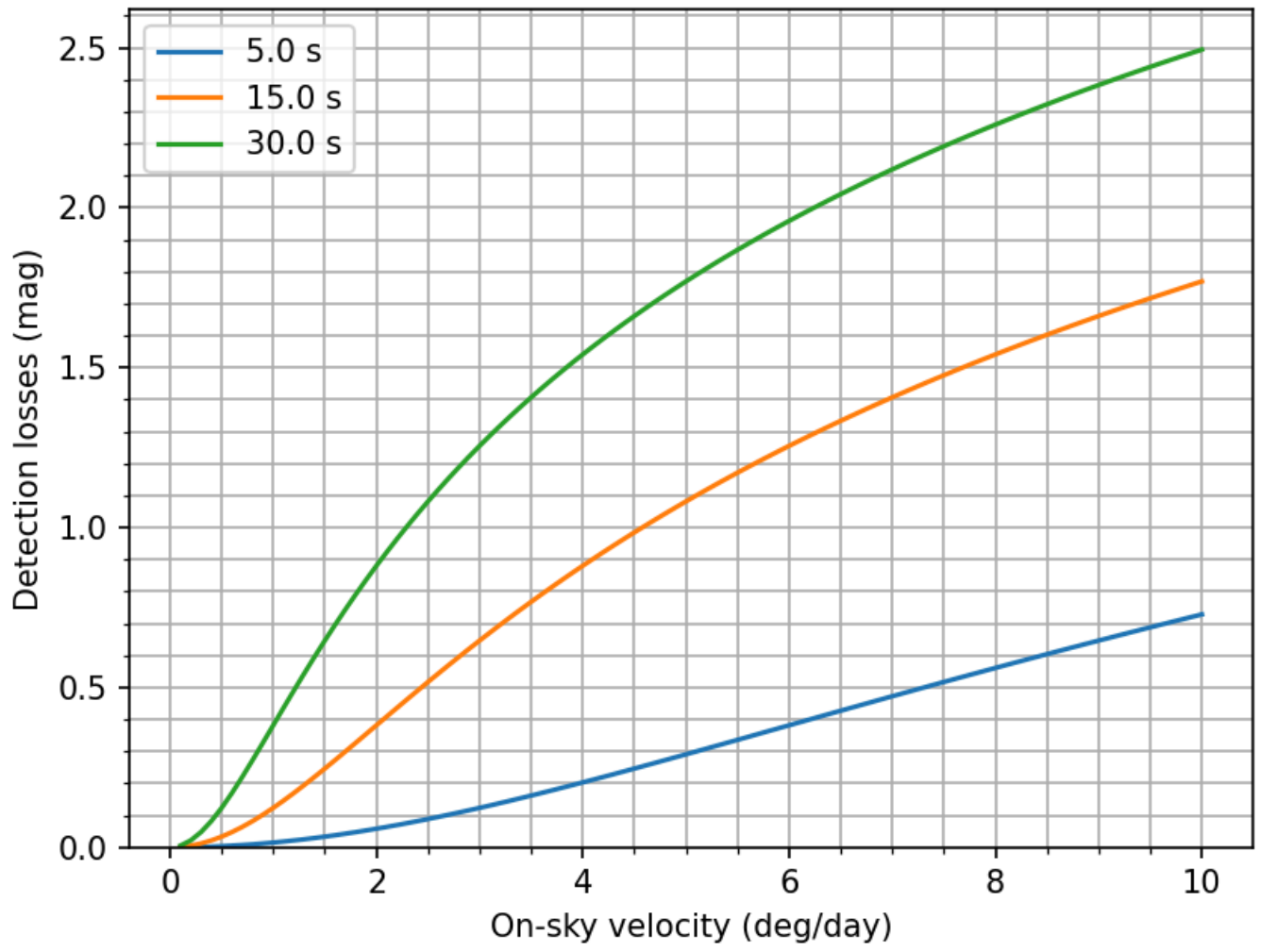

**Figure 9**: Decrease in detection depth due to motion-induced trailing losses (in magnitudes) for 30s (green, standard), 15s and 5s exposures as a function of on-sky velocity (in degrees/day). This assumes a typical $r$ filter FWHM of 0.8". Formula for trailing losses taken from `rubin_sim` and Jones et al. (2018).

## Rates and Target of Opportunity activation criteria

### Event rate (triggers/year)

Min: 1

Max: 30

Rates are very difficult to estimate since Rubin is so much larger and more sensitive than existing NEO surveys and the characteristics of the orbit quality and timescale of new close-passing NEO discoveries from Rubin (3 pairs of detections over a ~15-day period with each pair separated by 33 minutes producing a confirmed new object) than existing NEO surveys (3-4 images over ~40 minutes with confirmation of NEO nature & orbit refinement from other facilities) that prior precedent on rates is of limited use. This rate is based on the statistics of known close-approachers from the JPL Small-Body DB (SBDB) which is known to be biased low due to the absence of search assets for fast-moving small NEOs before Rubin. In addition, in order to be a close-approaching NEO and in the SBDB, they have to (by definition) be non-impacting. However, we have no access to the time-evolution of the impact probability as a function of time and amount of follow-up observations from the Scout alerts which is what we really need, in conjunction with a better estimate of the rate of small close approachers, to determine a more accurate rate of need for Rubin ToOs. This number will be better understood after Year 1 of LSST.

**Trigger format:** ToO triggers for close-approaching or potentially impacting "small PHAs" would occur based on the JPL Scout alerts[15] (for newly discovered/short arc objects from Rubin or other surveys) or Sentry alerts[16] (for objects with longer (multi-year) arcs that have updated orbits and newly increased impact probabilities) that have specific conditions; these conditions are listed below and would need to be extracted from the Scout alert information, which is in JSON format.

We recommend triggering a Rubin ToO for a close-approaching or potentially impacting "small PHA" under the following conditions:

### Trigger Criteria

- *H<28 (corresponding to diameter d>10m assuming albedo=0.15)*
- *Earth close approach distance < 1 Lunar distance (LD; limits Rubin ToO time to not trigger on non-threatening objects that are projected to pass by at much larger distance; also the rate of potential triggers increases greatly with increased radius)*
- *Cumulative impact probability > 10^-3*
- *Low likelihood that the object is in a geocentric orbit*
- *Predicted magnitude V>21.6 and ephemeris uncertainty >1 sq. deg (Northern Hemisphere) or V>21.8 and ephemeris uncertainty > 3 sq. deg (Southern Hemisphere) at T+24 hours from the time of the Scout/Sentry alert*
- *No community ToO follow-up programs exist for the current semester on community non-Rubin assets, attempts to recover with other facilities have not been successful, and searches for <5-sigma detections from any existing LSST observations have not been successful*
    - *If a follow-up program on a facility such as Gemini South or ESO VLT is granted time to observe a Rubin ToO target, the Rubin ToO could be paused; although this would be difficult to do programmatically and would likely require a human to be part of the trigger decision process*

[15] Table A2 and https://cneos.jpl.nasa.gov/scout/intro.html

[16] Table A3 and https://cneos.jpl.nasa.gov/sentr

- *Likewise, if the Rubin SSP team found the object in the <5-sigma Rubin detections, the Rubin ToO could be paused*
- *Object unlikely to be detected in the natural course of LSST observations*
- *Apparent sky-plane rates of <10 deg/day (This is an externally imposed restriction on Rubin processing for national security considerations ([DMTN-199](DMTN-199)) but excludes most targets that are within ~1 LD and within a few days of close-approach.)*
    - *Rubin's unprecedented depth leads to discoveries of close-passing NEOs at a much greater distance than e.g. 0.5-m ATLAS, allowing removal of the risk of impact at greater distances when it can't be followed up with smaller facilities. This circumvents the issue of trying to chase rapidly accelerating objects (which could exceed the Rubin 10 deg/day limit and have increased trailing losses) and that have growing uncertainty regions that could overwhelm smaller FoV facilities*

The above information would come from the JPL Scout alerts (for newly discovered/short arc objects) or Sentry alerts (for objects with longer (multi-year) arcs that have updated orbits and newly increased impact probabilities) that are updated and resent after the receipt and processing by the MPC and JPL of new observations from either Rubin or other follow-up facilities. The most recent alert will thus always contain the most up-to-date information, and therefore it is relatively easy to avoid a trigger or re-trigger of a Rubin ToO when an updated object would no longer pass the trigger criteria (primarily the impact probability and/or the ephemeris uncertainty) defined above. These alerts would need to be obtained through API calls developed by the Rubin Project to process and analyze the Scout and Sentry alerts and trigger Rubin ToOs programmatically. To ensure that the executed triggers are indeed valuable, we recommend that a human be kept in the ToO trigger loop, especially early on in science operations as we understand the false positive rate produced by early Rubin discoveries; the LSST Solar System Science Collaboration (SSSC) NEOs+ISOs Working Group (WG) can commit to having someone on call who can reply to emails from the observer before triggering a ToO on a close-passing/potentially impacting "small PHA" coming from a JPL or Sentry alert. The NEOs+ISOs WG would also monitor the trigger performance to assess metrics such as the overall rate of Rubin ToO triggers and the false positive rate.

## Time budget and observing strategy

Quantitative goals for the follow-up:

- **Depth:** close-approaching or impacting small PHAs will be brighter than at the discovery epochs so the regular visit exposure times/single visit depth will produce better SNR and therefore better astrometric precision than the original discovery observations

- **Time frame(s) relative to trigger:** within the same local night (or twilight) as the trigger to the end of the following night

- **Area:** enough visits to cover the predicted ephemeris uncertainty at the time of observing

- **Filters:** *r* band for the maximum sensitivity and highest SNR for obtaining the astrometry required to most rapidly improve the orbit determination and decrease the impact probability. As a 2nd order optimization to minimize filter changes, it would be acceptable to execute observations in either of the red-sensitive bands (*iz*) if that filter were in position at the time of ToO execution. This would be particularly beneficial during the limited observing time available within the near-Sun twilight microsurvey.

The close-approaching small PHAs from the twilight survey follow the same criteria and require the same follow-up as the regular close-approaching small PHAs case, so no trade-offs are necessary (beyond implementation details for Project on potentially needing to trigger ToOs during twilight time).

**Search Area (sq deg)**

**Min:** >5' (set by the smallest FoV of Gemini+GMOS-S and VLT+FORS2 for objects with V>22 which are too faint + rapidly moving to be recovered using 4m telescopes + 1-3 sq. deg imagers)

**Max:** few 10's of sq. degrees

**Typical:** < 1-5 degrees

**Max delay from trigger (hours)**

Ideally, observations commence within an hour of receipt of the ToO trigger. The uncertainty region and on-sky rate of motion will grow with increasing time delay from the trigger, leading to an increased chance of not being able to recover the object or link the observations to the existing object.

### Trigger time distribution or constraints

The trigger could occur within a few tens of minutes to a few days of initial detection or discovery by Rubin or another survey. The urgency is set by a combination of factors, namely:

- the predicted close approach distance and associated uncertainty on this particular close approach,
- the cumulative impact probability accumulated over current and future close approaches to the Earth,
- the predicted magnitude and size of the uncertainty region at T+24 hours after the alert trigger,
- the absolute magnitude and therefore the estimated diameter and impact energy were the object to potentially impact.

Close-approaching "small PHAs" are not known to have any preferred direction or occurrence time for objects of the size range under consideration. Objects detected in the near-sun twilight survey will only be able to be followed in twilight. Small PHAs discovered by Rubin to be on a potentially-impacting trajectory coming from the sunward direction found during twilight will require ToOs to be rapidly triggered and executed before that twilight period ends and the object is no longer visible in order to confirm or rule out the possibility of impact.

### Observing Strategy Details

Filter change time is ~2 minutes, slew & settle time is 10-15 s, readout is 2s.

The science case does not require observations after the first night of ToO activation (day 0). Although subsequent observations may be required to obtain orbit-improving astrometry and colors for characterization, these can be carried out as part of the regular Rubin survey or with other dedicated facilities and don't require Rubin ToOs.

**Day 0:**

For pure astrometry, *r*-band exposures are recommended. We recommend 3 epochs of observation centered on the predicted asteroid position at the time of the exposure. The epochs should be separated by ~33 minutes and include 2 exposures with a dither.

For astrometry to improve the orbit and reduce the probability of impact, we need detectable measurements from three to four epochs on the same night. Due to the active area fraction of 90% in the LSST Camera, we ask for the telescope pointing to be dithered between the two exposures in an epoch in order to reduce the risk of these fast-moving objects disappearing in an inactive area or chip gap.

number of Observations per filter: 6

**Total time:**

- Slew/settle, Setup and filter change: 15 sec (slew/settle; worst case) + 120 sec filter change = 135 sec setup (assuming no overlap of slew & filter change)
- 2x15 sec integrations + 2x2 sec (readout) + 3 sec (dither/settle) overheads = 37 sec per exposure
- Total = 135 sec + 6x37 sec = 357 sec total sequence time (0.099 hours)

Worst case scenario of 30 triggers/year: Total time = **2.98 hours/year**

This calculation above is for dark time. If the only time to observe the small PHA is in twilight, then the exposure should be reduced to 15 sec (or whatever is suitable for the sky background) using the *z*-band filter, like the near-Sun twilight microsurvey.

## Special Observing Requests and Data Processing Requirements

Because non-standard observations cannot be guaranteed and no special processing will be supported at the start of operations, we do not recommend exposure times != 30 s. Since Rubin's unprecedented depth will lead to discoveries of close-passing/potentially impacting small PHAs at a much greater distance from Earth than existing NEO survey facilities are able to accomplish, constrained as needed by Rubin ToOs, the risk of impact will be removed when objects are at greater distances from Earth, circumventing the issue of trying to follow-up rapidly accelerating objects that could lead to increased trailing losses. This should result in 30 s exposures that do not lead to significant trailing losses for Rubin ToOs. Additionally, the lack of support for rapid processing of exposure times != 30 s would make it difficult to execute a triggered ToO rapidly as would be required for small PHAs discovered by Rubin to be on a potentially-impacting trajectory coming from the sunward direction found during twilight that must be followed-up before that twilight period ends and the object is no longer visible in order to confirm or rule out the possibility of impact. Finally, using the standard 30 s exposure time will allow any Solar System ToOs executed during the regular nightly survey cadence to be folded back into the WFD observations, limiting the disruption and diminishing of the full survey caused by ToOs.

The standard DM products from the 2x15s exposures combined to single 30s exposure (or 1x30s if adopted) of accurate astrometric positions, precise exposure timing and accurate magnitudes, combined with the normal reporting channels to the MPC will be sufficient for this science case. In some situations, this may require some intervention by the Rubin SSP team to pull out the associated observations such as manual access to the pixel data inside the 80-hour limit or accessing the linked tracklets SSP has identified from the given night.

## Downscoping direction suggestions

As discussed above under the rates, the predicted rates of triggers are hard to estimate given the lack of precedent from existing surveys (opposite hemisphere, shallower depth, smaller FoV as discussed previously) but options for limiting the rate if it is judged to be excessive could include:

1) raising the threshold on impact probability, resulting in fewer triggers,
2) raising the absolute magnitude (*H*) cutoff to smaller values (larger object diameters) and therefore using ToO triggers on objects that have a higher risk of more significant damage,
3) increasing the time threshold after discovery but before triggering Rubin ToOs in the hope that other facilities that are observing at the same time i.e. Blanco+DECam/CFHT+Megacam (since there is a lack of deep+wide field capabilities at ESO or other Southern Hemisphere longitudes) are able to perform the initial recovery first

We also recommend that this ToO program only begin after the first 3 months of Rubin science operations in order to assess both the PHA impactor false positive rate and event rate when the influx of Rubin discoveries begin to populate the JPL Sentry Earth impact monitoring table.

## Total ToO time budget recommended for the PHA science case

| | |
|---|---|
| **PHA case** | **<30 hours** |

## Emerging Ideas Science Case: Future Interesting Twilight Discoveries

*Science case leads: Tim Lister, Sarah Greenstreet*

Rubin's near-Sun twilight microsurvey will offer the opportunity for Rubin to observe the low-SE sky during twilight, which is the only time when viewing Solar System objects inward to Earth is possible. Near-Earth objects (NEOs) interior to Earth's orbit (including Atiras with orbits interior to the orbit of Earth and 'Ayló'chaxnims (or inner-Venus asteroids) with orbits interior to Venus) are the least constrained portion of currently available NEO models owing to observational limitations of objects at low SE (Greenstreet et al., 2012; Granvik et al. 2018). In addition, the near-Sun twilight microsurvey could enhance the discovery of interstellar objects (ISOs); ISO 2I/Borisov was discovered during twilight by an amateur astronomer in 2019 (Borisov, 2019). Characterizing these interstellar visitors, including obtaining photometric colors, light curves & rotation periods, as well as possible mass shedding, outbursting, or breakup events provides a unique opportunity to help put our Solar System in context with other exoplanetary systems.

Some inner-Venus asteroids detected in Rubin's near-Sun twilight microsurvey could have large enough ephemeris uncertainties to become lost behind the Sun if not rapidly recovered by triggering a Rubin ToO during twilight hours. Likewise, some ISOs may have exceptionally limited observing windows/geometry before becoming too faint as they leave the Solar System that would require Rubin ToO follow-up before the unique opportunity to observe and characterize these interstellar visitors disappears.

However, due to a number of uncertainties, largely stemming from the current small sample sizes for these two populations (to date, one inner-Venus asteroid and two ISOs have been discovered), we have chosen not to pursue this science case in the current Rubin ToO proposal round. It's unclear whether the astrometric uncertainties for these population discoveries would warrant triggering Rubin ToOs, and thus the event rate for Rubin ToO triggers for these objects is also largely uncertain. The near-Sun twilight microsurvey should increase discoveries of both inner-Venus asteroids and ISOs, however, it will likely remain unclear how many objects may need rapid follow-up from Rubin and under what conditions that follow-up would need to be conducted until after the first year of the survey has been completed. At that time, we should have a better understanding of the discovery rates and the need to trigger Rubin ToOs for these populations. We may thus want the opportunity to recommend Rubin ToOs for this science case after Year 1 of the survey has been completed when the next round of Rubin ToO proposals are considered (Bianco et al., 2022).

# Appendix A

The SCIMMA MMA alert packets are described contain the following features that may be usedful in selecting targets:

| | | |
|---|---|---|
| **LKV alerts** | | |
| Packet Type | GCN Notice type: *{Preliminary,Initial,Update,Retraction}* | |
| FAR | Estimated False Alarm Rate | |
| Significant | 1 if trials factor × FAR < 1/month for CBC events, otherwise 0 | 1 if trials factor × FAR < 1/year for burst events, otherwise 0 |
| instruments | List of detectors, e.g. *['H1', 'L1','V1']* whose data have been used | |
| Sky Map | sky localization file (probability regions) | |
| Group | CBC | Burst |
| Pipeline | The pipeline that produced the alert | |
| central_frequency | estimation of the frequency of the signal's main component | |
| duration | time interval for which the signal is detected (above noise) | |
| BNS, NSBH, BBH, Noise | Probability that the source is a BNS, NSBH, BBH, or Terrestrial (i.e, noise) respectively | |
| HasNS, HasRemnant, HasMassGap | Probability, under the assumption that the source is not noise, that at least one of the compact objects was a neutron star, that the system ejected a non-zero amount of neutron star matter, and that at least one of the compact objects has mass in the range 3-5 solar masses, respectively | |
| WhereWhen | Time of signal | |
| **In the event of a coincidence between a gravitational-wave candidate and an alert from a third party (e.g. a gamma-ray burst or neutrino trigger), the following fields will also be present:** | | |
| External Observatory | [Fermi,Swift,INTEGRAL,AGILE,SNEWS] | |
| External Search | [GRB,SubGRB,SubGRBTargeted,Supernova] | |
| Time Coincidence FAR | Estimated coincidence false alarm rate in Hz using timing | |
| Time and Sky Position Coincidence FAR | Estimated coincidence false alarm rate in Hz using timing and sky position | |
| Joint Skymap Fits | combined GW-External sky localization file | |
| Time Difference | Time between source and external event in seconds | |

**Table A1:** Subset of features in the SCIMMA MMA alert packets (potentially) relevant to select trigger. For a complete list see https://emfollow.docs.ligo.org/userguide/content.html

| | |
|---|---|
| The number of observations and the length of the observed arc | |
| The RMS of the weighted residuals for the best fitting orbit | |
| Absolute magnitude *H* | A coarse indicators of, when defined |
| Close approach distance | A coarse indicators of, when defined |
| Minimum Orbital Intersection Distance (MOID) | A coarse indicators of, when defined |
| Velocity relative to Earth | A coarse indicators of, when defined |
| RA | Updated every 15 minutes: |
| DEC | Updated every 15 minutes: |
| *V*-band magnitude | Updated every 15 minutes: |
| Solar elongation | Updated every 15 minutes: |
| Plane-of-Sky (POS) rate of motion | Updated every 15 minutes: |
| 1-sigma POS uncertainty | Updated every 15 minutes: |
| POS uncertainty tomorrow | Projected, updated every 15 minutes: |
| Probabilistic object classification | [*PHA, NEO, geocentric, cometary (Tisserand parameter $T_J$ < 3), and Interior-to-Earth orbits*] Likelihood scores between 0 and 100 for |
| A rating to categorize the chances of an impact on Earth | [*Negligible, Small, Modest, Moderate, Elevated*] not defined for arcs shorter than 20 minutes or with less than three observations |

**Table A2:** Subset of features in the JPL Scout alert packets (potentially) relevant to select trigger. For a complete list see https://cneos.jpl.nasa.gov/scout/intro.html

| | |
|---|---|
| Impact Probability | Probability that the tabulated impact will occur. The probability computation is complex and depends on a number of assumptions that are difficult to verify. For these reasons the stated probability can easily be inaccurate by a factor of a few, and occasionally by a factor of ten or more. |
| Impact Energy (Mt) | Kinetic energy at impact, based upon the computed absolute magnitude and impact velocity for the particular case, and computed in accordance with the guidelines stated for the Palermo Technical Scale. Uncertainty in this value is dominated by mass uncertainty and the stated value will generally be good to within a factor of three. |
| Palermo Scale | Hazard rating according to the Palermo technical |
| | impact hazard scale, based on the tabulated impact date, impact probability and impact energy. |
| Torino Scale | Hazard rating according to the Torino impact hazard scale, based on the tabulated impact probability and impact energy. The Torino scale is defined only for potential impacts less than 100 years in the future. |

**Table A3:** Subset of features in the Sentry alert packets (potentially) relevant to select trigger. See https://cneos.jpl.nasa.gov/sentry/

## Acknowledgments

We are grateful to the Heising-Simons Foundation and U.C. Berkeley for financial and local support in the organization of the RubinToO24 workshop. We thank the members of the Vera C. Rubin Observatory who participated in the workshop and who, while not authors of this report, provided invaluable expertise and information indispensable to the formulation of this document: Robert Blum, Zeljko Ivezic, Steven Ritz, Steven Kahn. We further thank all participants in the workshop. Although some did not contribute to the text directly and have opted not to be included as authors, in many cases, their contributions and support to the workshop discussions substantially shaped the direction and content submitted.

# GLOSSARY OF ACRONYMS AND TERMS

A:

AB: the AB magnitude system

AGN: Active Galactic Nuclei

AGILE: Astro-rivelatore Gamma a Immagini Leggero

AP: Alert Production (team of Rubin Observatory)

API: Application Programming Interface

ASAP: As Soon As Possible

ASAS-SN: All-Sky Automated Survey for Supernovae

AT: Astronomical Transient

ATLAS: Asteroid Terrestrial-impact Last Alert System

B:

BBH: Black Bole-Black Hole (merger)

BGP: Binned Gaussian Probability

BNS: Binary Neutron Star (merger)

C:

CCSN: Core collapse Supernova

CFHT: Canada France Hawaii Telescope

CL: Confidence Limit

CBC: Compact Binary Coalescence

ComCam: The Rubin Observatory Commissioning Camera

D:

DB: Data Base

DEC: Declination

DECam: Dark Energy Camera

DESC: Dark Energy Science Collaboration: one of the 8 Rubin LSST Science Collaborations

DIA: Difference Image Analysis

DM: Data Management (team of Rubin Observatory)

DMTN: Data Management Technical Note (of Rubin Observatory)

DUNE: Deep Underground Neutrino Experiment

E:

EM: Electromagnetic

ESO: European Space Organization

F:

FAR: False Alarm Rate

FBOT: Fast Blue Optical Transient

FORS2: FOcal Reducer/low dispersion Spectrograph 2

FoV: Field of View

FWHM: Full Width Half Maximum

G:

GCN: General Coordinates Network

GMOS: Gemini Multi-Object Spectrographs

GR: General Relativity

GRB: Gamma Ray Burst

GVD: Gigaton Volume Detector

GW: Gravitational Wave

HLVK

I:

IGWN: International Gravitational-Wave Observatory Network

INTEGRAL: INTErnational Gamma-Ray Astrophysics Laboratory

ISO: Inter-Stellar Object

J:

JPL: NASA Jet Propulsion Laboratory

JSON: JavaScript Object Notation

JWST: JW Space Telescope

K:

KM3NeT: Cubic Kilometre Neutrino Telescope

KN: Kilonova

L:

LD: Lunar Distance

LFBOT: Luminous Fast Blue Optical Transient

LIGO: Laser Interferometer Gravitational-wave Observatory

LSST: Legacy Survey of Space and Time

LSSTCam: The Rubin Observatory survey camera

LSST:UK: the LSST UK consortium

LVK: LIGO-Virgo-KAGRA

M:

MITSuME: Multicolor Imaging Telescopes for Survey and Monstrous Explosions

MMA: Multi-Messenger Astronomy

MOID: Minimum Orbital Intersection Distance

MPC: Minor Planet Center

N:

NEO: near-Earth object

NGC: New General Catalogue of Nebulae and Clusters of Stars (Dreyer 1888)

NIR: Near InfraRed

NS-BH: Neutro Star-Black Hole (merger)

O:

O#: number (#) of the LVK observing run, e.g. O5 is Observing Run 5

P:

Pan-STARRS: Panoramic Survey Telescope and Rapid Response System

PHA: Potentially Hazardous Asteroid

POS: Plane-of-Sky

PSTN: Project Science Technical Note  (of the Rubin Observatory)

R:

RA: Right Ascention

$R_{BNS}$: rate of BNS mergers

S:

SBDB: Small Bodies DataBase

SBO: Shock Breakout

SCiMMA: Scalable Cyberinfrastructure to support Multi-Messenger Astrophysics

SCOC: Survey Cadence Optimization Committee

SE: Solar Elongation

SFR: Star Formation Rate

SLSC: Strong Lensing Science Collaboration: one of the 8 Rubin LSST Science Collaborations

SLTT: Strong Lensing Topical Team (within the DESC)

SN: Supernova

SNEWS: Supernova Early Warning System

SNR: Signal-to-Noise Ratio

SSSC: Solar System Science Collaboration: one of the 8 Rubin LSST Science Collaborations

Super-K: Super Kamiokande (observatory)

T:

TBD: To Be Determined

TDE: Tidal Disruption Event

TJ: Tisserand parameter with respect to Jupiter

ToO: Target of Opportunity

TVS: Transient and Variable Stars (Science Collaboration): one of the 8 Rubin LSST Science Collaborations

TXS: Texas Survey (of radio sources)

U:

UVOT: Ultra-violet Optical Telescope (at Swift)

V:

VLT: Very Large Telescope

W:

WFD: Wide Fast Deep - the main LSST survey

WG: Working Group

Z:

ZTF: Zwicky Transient Facility

# AUTHOR AFFILIATIONS

(1) Department of Physics and Astronomy, University of North Carolina at Chapel Hill, Chapel Hill, NC 27599-3255, USA
(2) Joint Space-Science Institute, University of Maryland, College Park, MD 20742, USA
(3) Department of Astronomy, University of Maryland, College Park, MD 20742, USA
(4) Astrophysics Science Division, NASA Goddard Space Flight Center, Mail Code 661, Greenbelt, MD 20771, USA
(5) Department of Astronomy, University of California, Berkeley, CA 94720, USA
(6) Department of Physics, University of California, 366 Physics North MC 7300, Berkeley, CA 94720, USA
(7) Brookhaven National Laboratory, Bldg 510, Upton, NY 11973, USA
(8) Vera C. Rubin Observatory/NSF NOIRLab 950 N Cherry Ave., Tucson, AZ, 85749, USA
(9) Department of Astronomy and the DiRAC Institute, University of Washington, Seattle, WA, USA
(10) Las Cumbres Observatory, 6740 Cortona Drive Ste 102, Goleta, CA, 93117, USA

(11) McWilliams Center for Cosmology and Astrophysics, Department of Physics, Carnegie Mellon University, Pittsburgh, PA 15213, USA
(12) INAF, Osservatorio Astronomico di Roma, via Frascati 33, I-00078 Monte Porzio Catone (RM), Italy
(13) Department of Physics, University of Oxford, Keble Road, Oxford, OX1 3RH, UK
(14) Astrophysics Research Centre, School of Mathematics and Physics, Queen's University Belfast, BT7 1NN, UK
(15) School of Physics and Astronomy, University of Birmingham, Birmingham B15 2TT, UK
(16) Division of Physics, Mathematics, and Astronomy, California Institute of Technology, Pasadena, CA 91125, USA
(17) Division of Physics, Mathematics and Astronomy, California Institute of Technology, Pasadena, CA 91125, USA
(18) Department of Physics, Stanford University, 382 Via Pueblo Mall, Stanford, California 94305, USA
(19) Kavli Institute for Particle Astrophysics & Cosmology, P. O. Box 2450, Stanford University, Stanford, California 94305, USA
(20) SLAC National Accelerator Laboratory, Stanford University, 2575 Sand Hill Road, Menlo Park, CA 94025, USA
(21) OzGrav, School of Physics, The University of Melbourne, Parkville, VIC 3010, Australia
(22) Department of Astronomy and Astrophysics, University of California, Santa Cruz, CA, 95064, USA
(23) School of Physical and Chemical Sciences|Te Kura Matu, University of Canterbury, Private Bag 4800, Christchurch 8140, New Zealand
(24) Lawrence Berkeley National Laboratory, 1 Cyclotron Road, MS 50B-4206, Berkeley, CA 94720, USA
(25) Eureka Scientific, 2452 Delmer Street, Oakland, CA 94602
(26) Centro Brasileiro de Pesquisas F´ısicas, Rua Dr. Xavier Sigaud 150, Rio de Janeiro, 22290-180, RJ, Brazil
(27) Centro Federal de Educacao Tecnologica Celso Suckow da Fonseca, Rodovia Marcio Covas, lote J2, quadra J, Itaguai, Brazil
(28) University of Arizona, Lunar and Planetary Laboratory, 1629 E. University Blvd.,Tucson, AZ 85721, USA
(29) South African Astronomical Observatory, PO Box 9, Observatory 7935, Cape Town, South Africa
(30) Department of Astronomy, University of Cape Town, Private Bag X3, Rondebosch 7701, South Africa
(31) Department of Physics, University of the Free State, P.O. Box 339, Bloemfontein 9300, South Africa
(32) National Radio Astronomy Observatory, 520 Edgemont Rd, Charlottesville VA 22903, USA

(33) Centre for Astrophysics and Supercomputing, Swinburne University of Technology, Hawthorn, VIC 3122, Australia
(34) Australian Research Council Centre of Excellence for Gravitational Wave Discovery (OzGrav), Australia
(35) Department of Physics and Astronomy, Texas Tech University, Lubbock, TX, United States
(36) School of Physics and Astronomy, University of Minnesota, Minneapolis, Minnesota 55455, USA
(37) Benemérita Universidad Autónoma de Puebla, Av. San Manuel, 72000 Puebla, Mexico
(38) INAF-IRA, Via P. Gobetti 101, I-40129, Bologna, Italy
(39) Institute of Astronomy and Kavli Institute for Cosmology, University of Cambridge, Madingley Road, Cambridge CB3 0HA, UK
(40) Ruhr University Bochum, Faculty of Physics and Astronomy, Astronomical Institute (AIRUB), Universitätsstraße 150, 44801 Bochum, Germany
(41) Center for Astrophysics and Cosmology, University of Nova Gorica, Vipavska 13, 5000 Nova Gorica, Slovenia
(42) Institute for Gravitational Wave Astronomy, University of Birmingham, Birmingham B15 2TT, UK
(43) Gran Sasso Science Institute (GSSI) L'Aquila, Italy
(44) INAF-Astronomical Observatory of Abruzzo, Via Maggini snc, 64020, Teramo, Italy
(45) INFN Laboratori Nazionali del Gran Sasso, L'Aquila, Italy
(46) Department of Physics and Astronomy, University of Pittsburgh, 3941 O'Hara St, Pittsburgh, PA 15260
(47) Department of Astronomy & Astrophysics, University of California, San Diego, 9500 Gilman Drive, MC 0424, La Jolla, CA 92093-0424, USA
(48) Center for Astrophysics | Harvard & Smithsonian, 60 Garden Street, Cambridge, MA 02138-1516, USA
(49) INAF, Osservatorio Astronomico di Capodimonte, Salita Moiariello 16, I-80131 Napoli, Italy
(50) DARK, Niels Bohr Institute, University of Copenhagen, Jagtvej 128, 2200 Copenhagen, Denmark
(51) United States Naval Academy
(52) Astrophysics Research Institute, Liverpool John Moores University, IC2 Liverpool Science Park, 146 Brownlow Hill, Liverpool, L3 5RF, UK
(53) Department of Astrophysics, Institute for Mathematics, Astrophysics and Particle Physics (IMAPP), Radboud University, Nijmegen, The Netherlands
(54) Department of Physics, University of Warwick, Coventry, UK
(55) Physik-Institut, University of Zurich, Winterthurerstrasse 190, 8057 Zurich, Switzerland
(56) Campo Catino Astronomical Observatory, Regione Lazio, Guarcino (FR), 03010, Italy
(57) McWilliams Center for Cosmology and Astrophysics, Department of Physics, Carnegie Mellon University, Pittsburgh, PA, USA

(58) Department of Physics and Astronomy, Northwestern University, 2145 Sheridan Rd, Evanston, IL 60208, USA
(59) Center for Interdisciplinary Exploration and Research in Astrophysics (CIERA), Northwestern University, 1800 Sherman Ave, Evanston, IL 60201, USA
(60) EXPOASTRONOMY Astrophysics and Space Sciences Research Group, Global Sky Partner of Las Cumbres Observatory
(61) University of Chile
(62) Inter-University Centre for Astronomy and Astrophysics (IUCAA)
(63) University of Illinois, Urbana-Champaign, Dept. of Astronomy, 1002 W. Green St., Urbana, IL 61801, USA
(64) Center for AstroPhysical Surveys, NCSA, 1205 W Clark St, Urbana, IL 61801, USA
(65) Astrophysics Research Centre, School of Mathematics and Physics, Queens University Belfast, Belfast BT7 1NN, UK
(66) Princeton University, Department of Astrophysical Sciences
(67) Legacy Survey of Space and Time Discovery Alliance Catalyst Fellow
(68) Princeton University, Department of History
(69) Department of Physics, Lancaster University, Lancaster, LA1 4YB, UK
(70) Center for Machine Intelligence and Data science (C-MInDS), IIT Bombay, India
(71) Dipartimento di Fisica “E. Pancini”, Università di Napoli “Federico II”, Via Cinthia 21, I-80126 Napoli, Italy
(72) INAF – Osservatorio Astronomico di Capodimonte, Salita Moiariello 16, I-80131 Napoli, Italy
(73) Institute of Cosmology and Gravitation, University of Portsmouth, Burnaby Rd, Portsmouth, PO1 3FX, UK
(74) Oskar Klein Centre for Cosmoparticle Physics, Department of Physics, Stockholm University, AlbaNova, Stockholm SE-106 91, Sweden
(75) Nordita, Stockholm University and KTH Royal Institute of Technology, Hannes Alfvéns väg 12, SE-106 91 Stockholm, Sweden
(76) Astrophysics Research Centre, School of Mathematics and Physics, Queen's University Belfast, Belfast BT7 1NN, UK
(77) Department of Physics and Astronomy, Rutgers, the State University of New Jersey, 136 Frelinghuysen Road, Piscataway, NJ 08854-8019, USA
(78) Department of Astronomy and Carl Sagan Institute, Cornell University, 122 Sciences Drive, Ithaca, NY, 14853, USA
(79) NSF Astronomy and Astrophysics Postdoctoral Fellow
(80) Institut für Physik und Astronomie, Universität Potsdam, Haus 28, Karl-Liebknecht-Str. 24/25, 14476 Potsdam, Germany.
(81) Steward Observatory, University of Arizona, 933 North Cherry Avenue, Tucson, AZ 85721-0065, USA
(82) Department of Physics, University of Michigan, Ann Arbor, MI 48109 USA

(83) INAF-Istituto di Astrofisica e Planetologia Spaziali, via Fosso del Cavaliere, 100, I-00133 Rome RM, Italy
(84) International Centre for Space and Cosmology, School of Arts and Sciences, Ahmedabad University, Navrangpura, Ahmedabad, 380009, India
(85) Department of Physics and Astronomy, University of California, One Shields Avenue, Davis, CA 95616, USA
(86) SLAC National Accelerator Laboratory, 2575 Sand Hill Road, Menlo Park, CA 94025, USA
(87) Sub-department of Astrophysics, Denys Wilkinson Building, University of Oxford, Keble Road, Oxford, OX1 3RH, U.K.
(88) The NSF AI Institute for Artificial Intelligence and Fundamental Interactions
(89) Planetary Science Institute, 1700 East Fort Lowell, Suite 106, Tucson, AZ 85719, USA
(90) Department of Astronomy, University of Texas at Austin, Austin, TX, USA
(91) University of Delaware Department of Physics and Astronomy 217 Sharp Lab Newark, DE 19716 USA
(92) University of Delaware Joseph R. Biden, Jr. School of Public Policy and Administration, 184 Academy St, Newark, DE 19716 USA
(93) University of Delaware Data Science Institute
(94) Center for Urban Science and Progress, New York University, 370 Jay St, Brooklyn, NY 11201, USA